\documentclass[useAMS,usenatbib]{mn2e}
\usepackage{epsfig}
\usepackage{multirow}
\usepackage{natbib}
\usepackage{hyperref}
\usepackage{caption}
\usepackage{subcaption}
\usepackage{comment}
\usepackage{mathrsfs}
\usepackage{amsmath}
\usepackage{amssymb}
\usepackage{graphicx}
\providecommand{\e}[1]{\ensuremath{\times 10^{#1}}}
\title[kHz QPOs and Fe lines in the LMXB 4U 1636--53]{Broad iron emission line and kilohertz quasi-periodic oscillations in the neutron star system 4U 1636--53}

\author[Sanna et al. ]{Andrea Sanna$^{1,2}$\thanks{E-mail:
    andrea.sanna@dsf.unica.it}, Mariano M\'{e}ndez$^{1}$, Diego Altamirano$^{3,4}$, Tomaso Belloni$^{5}$, \newauthor
    Beike Hiemstra$^{1}$, Manuel Linares$^{6,7}$\\
$^{1}$Kapteyn Astronomical Institute, University of Groningen, PO Box 800, 9700 AV   Groningen, The Netherlands.\\ 
$^{2}$Dipartimento di Fisica, Universit\`a degli Studi di Cagliari, SP Monserrato-Sestu km 0.7, 09042 Monserrato, Italy \\ 
  $^{3}$Astronomical Institute, ``Anton Pannekoek'', University of Amsterdam,\\
  Science Park 904, 1098 XH Amsterdam, The Netherlands.\\ 
  $^{4}$Physics \& Astronomy, University of Southampton, Southampton, Hampshire SO17 1BJ, UK\\
$^{5}$INAF -- Osservatorio Astronomico di Brera, Via E. Bianchi 46, I-23807, Merate (LC), Italy\\
$^{6}$Instituto de Astrof{\'i}sica de Canarias, V\'ia L\'actea,  E-38205 La Laguna, Tenerife, Spain\\
$^{7}$Universidad de La Laguna, Dept. Astrof{\'i}sica, E-38206 La Laguna, Tenerife, Spain}

\begin{document}

\date{Accepted 2014 March 11.  Received 2014 March 10; in original form 2014 January 21 }

\pagerange{\pageref{firstpage}--\pageref{lastpage}} \pubyear{2002}

\maketitle

\label{firstpage}

\begin{abstract}

  Both the broad iron (Fe) line and the frequency of the kilohertz quasi-periodic oscillations
  (kHz QPOs) in neutron star low-mass X-ray binaries (LMXBs) can potentially provide independent
  measures of the inner radius of the accretion disc. We use \textit{XMM-Newton} and simultaneous \textit{Rossi X-ray Timing Explorer} observations
  of the LMXB 4U~1636--53 to test this hypothesis. 
  We study the properties of the Fe-K$\alpha$ emission line as a
  function of the spectral state of the source and the
  frequency of the kHz QPOs. We find that the inner radius of the
  accretion disc deduced from the frequency of the upper kHz QPO varies as a
   function of the position of the source in the colour-colour diagram, in accordance with previous work and with the 
standard scenario of accretion disc geometry. On the contrary, the inner disc radius
deduced from the profile of the iron line is not correlated with the spectral state of the source.
 The values of the inner radius inferred from kHz QPOs
  and iron lines, in four observations, do not lead to a consistent value
 of the neutron star mass, regardless of the model used to fit the iron line.
Our results suggest that either the kHz QPO or the standard relativistic Fe line
interpretation does not apply for this system. 
  Furthermore, the simultaneous detection of kHz QPOs and
broad iron lines is difficult to reconcile with models in which the broadening of the
iron line is due to the reprocessing of photons in an outflowing wind.
\end{abstract}


\section{Introduction}

The energy and power density spectra of low-mass X-ray binaries
(LMXBs) change with time in a correlated way, generally following
changes of the source luminosity, supporting the scenario in which 
these changes are a function of mass accretion rate in the system 
\citep[e.g., ][]{Wijnands97, Mendez99, GierlinskiDone02}. Evolution of the broad-band energy spectrum
in low-luminosity systems is thought to reflect changes in the configuration 
of the accretion-disc flow \citep[see review by][and references therein]{Done07}.
\citet{GierlinskiDone02} find a strong correlation in the LMXB 4U 1608--52 between 
the position of the source in the colour-colour diagram and the truncation radius of the 
inner accretion disc, which is likely driven by the average 
mass accretion rate through the disc. At low luminosity, the spectrum is consistent with emission
from an accretion disc truncated far from the neutron star; as the luminosity increases the 
spectrum softens and the inner radius of the accretion disc moves inwards.

\noindent
In a very similar way, changes of the power density spectra appear to be driven by 
mass accretion rate. At low luminosity, when the energy spectrum of the source is hard, all timing 
components in the power spectrum have relatively low characteristic frequencies. These 
frequencies increase as the energy spectrum softens and the inferred mass accretion rate through the disc increases \citep[see e.g.,][]{vdKlis97,
Mendez97,Mendez99,Mendez01, Homan02,Straaten02,Straaten03,Straaten05,Altamirano05,
Altamirano08a, Altamirano08b, Linares05,Linares07}. The fact that fits to
the energy spectra suggest that the accretion disc moves closer to the NS, and that the characteristic 
frequencies in the power density spectra increase as the luminosity increases, 
supports the idea that those frequencies are set by the dynamical frequencies in the accretion disc. The kilohertz quasi-periodic oscillations (kHz QPOs) are 
especially interesting because of the close correspondence between their frequencies and the Keplerian frequency
at the inner edge of the accretion disc \citep[e.g.,][]{MillerLambPsaltis98,Stella98}. On short time-scales (within a day or less),
the frequency of the kHz QPOs increases monotonically as the source brightens, and the inferred mass accretion
rate increases. However, on longer time-scales this correlation breaks down and the intensity-frequency 
diagram shows the so-called ``parallel tracks'' \citep{Mendez99}.
 
 Broad asymmetric iron (Fe) lines have been often observed in accreting systems with the compact object
 spanning a large range of masses, from supermassive black holes in AGNs \citep[see][for an extensive review]{Fabian00}
 to stellar-mass black holes \citep[e.g.,][]{Miller02,Miller04} and neutron star systems \citep{Bhattacharyya07}.
 The Fe K-$\alpha$ emission line at 6--7 keV is an important feature of the spectrum that 
 emerges from the accretion disc as a result of reflection of the corona and the NS surface/boundary layer photons
 off the accretion disc.
 The mechanism responsible for the broad asymmetric profile of the line is still under discussion. 
 \citet{Fabian89} proposed that the line is broadened by Doppler and relativistic effects due to motion of the matter in the accretion disc. \citet{diSalvo05} and \citet{Bhattacharyya07} discovered broad iron lines in the NS LMXBs
 4U 1705--44 and Serpens X-1, respectively. \citet{Cackett08} confirmed \citet{Bhattacharyya07} results using  
 independent observations, and also discovered broad, asymmetric, Fe K-$\alpha$ emission lines
 in the LMXBs 4U 1820--30 and GX 349+2. All these authors interpreted the broadening of the line
 as due to relativistic effects. Relativistic Fe lines have been observed at least in a dozen NS binary systems
 in the last decade \citep{diSalvo05,Bhattacharyya07,Cackett08, Pandel08, Cackett09b, Papitto09,diSalvo09,
 Reis09, Iaria09,DAi09,Cackett10,DAi10,Egron11,Cackett12,Sanna13}. A different interpretation for the broadening of
 the line has been suggested by \citet{Ng10} who re-analysed the data of several NS systems showing Fe K-$\alpha$ 
 emission lines. \citet{Ng10} claim that for most of the lines there is no need to invoke special and general
 relativity to explain the broad profile, and that Compton broadening is enough.
 
 If the relativistic interpretation of the Fe line is correct, we can directly test
 accretion disc models by studying the line properties, as the shape of the profile depends on the inner 
 and the outer disc radius. We expect then the iron line to be broader in the soft state -- when the inner radius 
 is smaller and the relativistic effects stronger -- than in the hard state.
 
 Accretion disc models can also be tested using simultaneous measurements of kHz QPOs and iron lines 
 \citep{Piraino00,Cackett10}. If both the Keplerian interpretation of the kHz QPOs frequency and the 
 broadening mechanism (Doppler/relativistic) of the Fe line are correct, these two observables
 should provide consistent information about the accretion disc. Furthermore, if the changes in the spectral continuum also
 reflect changes of the inner edge of the accretion disc, the Fe line should vary in correlation with the
 frequency of the kHz QPOs \citep[see, e.g.,][]{Bhattacharyya07}. Understanding the relation between kHz QPOs,
 Fe emission line and spectral states may have an impact beyond accretion disc physics. As discussed by \citet{Piraino00},
 \citet{Bhattacharyya07}, and \citet{Cackett08}, measurements of the line could also help constraining the mass and radius of the neutron star \citep{Piraino00}, parameters needed to determine the neutron star equation of state. \citet{Cackett10}\footnote{At about the same time, Altamirano et al. (2010) presented a similar analysis in a manuscript that was eventually never published (c.f. \S 5 in  \citet{Cackett10})} tested this idea using three observations of 4U 1636--53.
 
  With all of this in mind, in 
 this paper we investigate the correlation between the iron line, kHz QPOs and spectral states in the
 LMXB 4U 1636--53, with the aim of understanding whether the existing interpretations of these phenomena 
 are consistent. 
 
 The fact that 4U 1636--53 is well sampled with \textit{XMM-Newton}, and \textit{Rossi X-ray Timing Explorer (RXTE)} observations, is one 
 of the most prolific sources of kHz QPOs, shows strong Fe-K$\alpha$ lines, and shows regular 
 hard-to-soft-to-hard state transitions on time scales of weeks, makes this source an excellent target for this study.

\section{Observations}
\label{sec:obs}
The NS LMXB 4U 1636--53 has been observed with \textit{XMM-Newton} nine times between 2000 and
2009. The first 2 observations (in 2000 and 2001) were performed in imaging-mode, and suffered from severe 
pile-up and data loss due to telemetry drop outs; we therefore excluded these two observations from the analysis. In \citet{Sanna13}
we reanalysed the 2005, 2007 and 2008 observations \citep{Pandel08,Cackett10} and we analysed for the first time four new observations taken in 2009. 
The observation taken on March 14th, 2009 had a flaring high-energy background during the full 
$\sim$40 ks exposure, and we therefore excluded this observation from the analysis \citep[see][for more details
on the analysis of the \textit{XMM-Newton} observations]{Sanna13}. In this paper we make use of the results reported there.

Since 1996, 4U 1636--53 was observed $\sim$1300 times with \textit{RXTE}; from March 2005 the source was regularly observed
for $\sim$2 ks every other day (except for periods with solar viewing occultation). The first results of this monitoring
campaign have been reported by \citet{Belloni07}, \citet{Zhang12}, and \citet{Sanna12a}. We took all the information related to the kHz QPOs used for this paper from \citet[][]{Sanna12a}. Following \citet{Sanna13}, we refer to the 2005, 2006, 2007,
25th March 2009, 5th September 2009, and 11th September 2009 \textit{XMM-Newton} observations as Obs.~1--6,
 respectively. We will use the same labelling (Obs.~1, etc.) for the simultaneous \textit{RXTE} observations used to study the 
X-ray variability. When these names are used in the context of power spectral 
components, spectral hardness, intensities and colours, we refer to the \textit{RXTE} observations.

\section{Data analysis}
\subsection{Timing analysis}
We used data from 1280 \textit{RXTE} observations with the Proportional Counter Array \citep[PCA,][]{Zhang93}, covering
14 years of data since 1996. We produced light curves and colours using the Standard 2 data as in 
\citet{Altamirano08a}: we used 16-s time-resolution Standard 2 data to calculate the hard and soft colours, defined
as the 9.7--16.0 keV / 6.0--9.7 keV and 3.5--6.0 keV / 2.0--3.5 keV count rate ratio, respectively. We measured the intensity in
 the 2.0--16.0 keV range. We normalised the colours and the intensity by the corresponding Crab Nebula
values \citep[closest in time and within the same PCA gain epoch; e.g.,][]{Kuulkers94,Straaten03}.

We produced Fourier power density spectra (PDS) using the 2--60 keV data from the $\sim$125 $\mu$s (1/8192 s)
time resolution Event mode. We created Leahy-normalised PDS for each 16-s data segment with time bins of
1/8192 s, such that the lowest available frequency is 1/16 Hz, and the Nyquist frequency is 4096 Hz. We fitted the PDS
in the 50--4000 Hz range using a combination of one or two Lorentzians to fit the kHz QPOs, a constant for the 
Poissonian noise, and if needed, an extra Lorentzian to fit the broad-band noise at frequencies between 50 Hz and the
frequency range spanned by the kHz QPOs.

\begin{figure*}

\resizebox{2\columnwidth}{!}{\rotatebox{0}{\includegraphics[clip]{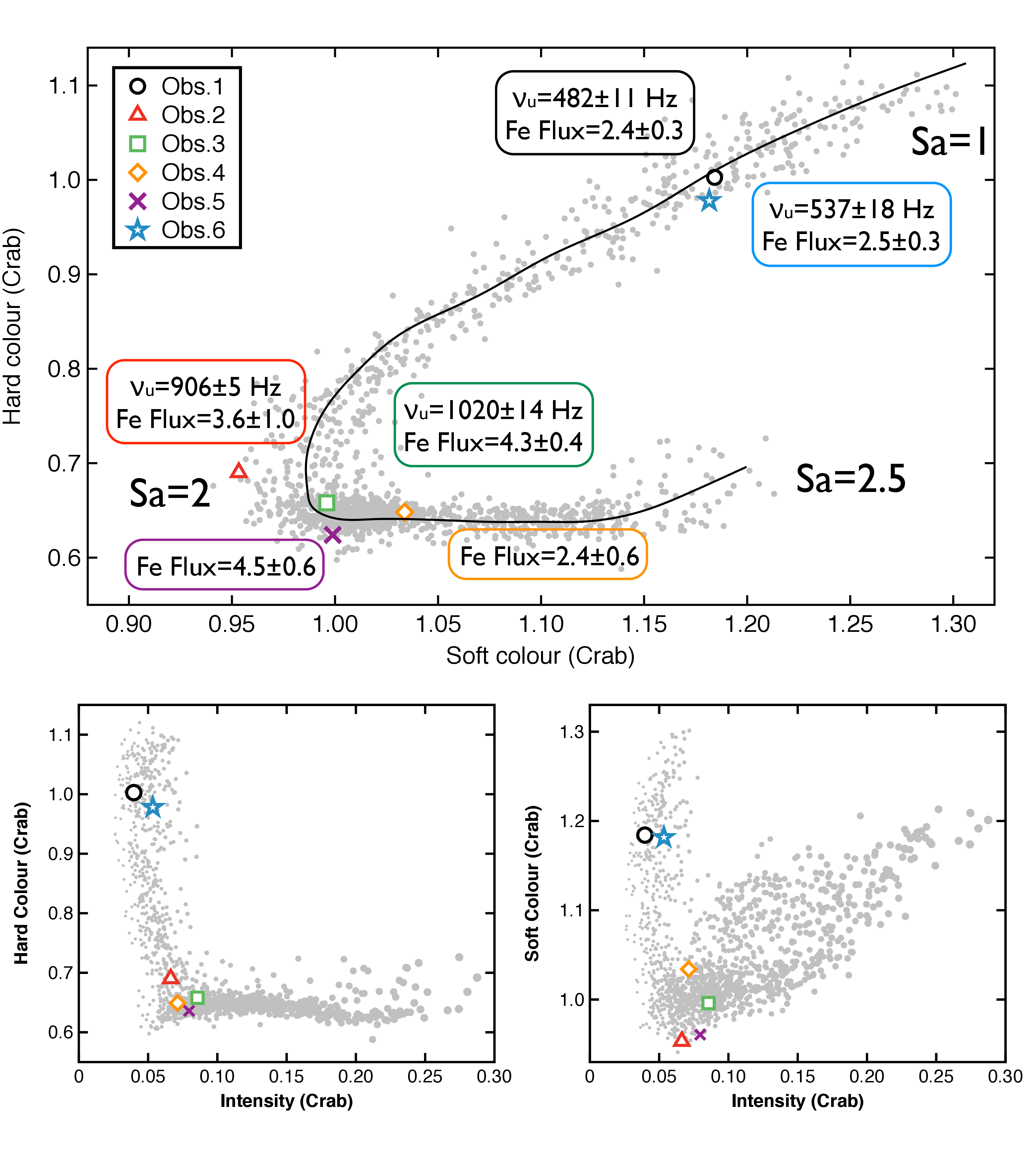}}}

\caption{Colour-colour, hard colour vs. intensity and soft colour vs. intensity
  diagrams (upper, lower left and lower right panels, respectively) of
  4U~1636--53 for all available \textit{RXTE} pointed observations (grey
  circles).
Colours and intensities during \textit{XMM-Newton} observations 
computed from pointed 
\textit{RXTE} observations (see Section~\ref{sec:obs})
are marked with different symbols following the legend at the upper-left corner. The position of the source on the colour-colour diagram is parametrized by the length of the black solid line that represents the coordinate $S_a$.
For each observation we report the frequency of the upper kHz QPO (when detected) 
and the flux of the line (Fe Flux) in units of $10^{-3}$ photons cm$^{-2}$ s$^{-1}$ for fits with a \textsc{kyrline} model with $a_*=0.27$ \citep[see][for more details]{Sanna13}.
}
\label{fig:cc}
\end{figure*}
\subsection{Spectral analysis}
\label{models}
For this work we use the results of the spectral analysis by \citet{Sanna13}, that we briefly summarised hereunder.\\
The energy spectrum of 4U 1636-53 can be well fitted with a multicolour disc blackbody (\textsc{DISKBB}) plus a single-temperature blackbody (\textsc{BBODY}) and a thermally comptonized component (\textsc{NTHCOMP}) to account for the thermal emission from the accretion disc, the thermal emission from the NS surface/boundary layer, and the high-energy emission from the corona-like region surrounding the systems, respectively. The soft seed photons of the comptonised component were assumed to come from the accretion disc. Besides the continuum emission, the data require a emission-like feature to model prominent residuals in the energy range 4-9 keV, around the Fe-K$\alpha$ emission line region. \citet{Sanna13} fitted this feature with a set of so-called phenomenological models (\textsc{gaussian, diskline, laor}, and \textsc{kyrline}), that only model the Fe emission line, and with two reflection models (\textsc{rfxconv}, and  \textsc{bbrefl}), which self-consistently model the whole reflection spectrum, of which the Fe-K$\alpha$ emission line is the strongest feature. \citet{Sanna13} showed that the Fe-K$\alpha$ emission line is well fitted by a symmetric gaussian profile characterised, however, by a large breadth, which is difficult to explain with mechanisms other than relativistic broadening.

\section{Results}

\subsection{Long term spectral behaviour of 4U 1636--53}

In Figure~\ref{fig:cc} we show the colour-colour (top), hard colour vs. intensity (bottom left), and the soft colour vs. intensity (bottom right)
diagrams for all the available \textit{RXTE} pointed observations. We use different symbols and colours to mark the location of the source 
during the \textit{XMM-Newton} observations as estimated by the simultaneous \textit{RXTE} observations (see Section~\ref{sec:obs}). 

Although 4U 1636--53 is a persistent X-ray source, it shows variations in intensity up to a factor of 6, following a narrow track
in the colour-colour and colour-intensity diagrams \citep{Belloni07,Altamirano08a}. Using data from the all-sky monitor (ASM)
onboard \textit{RXTE}, \citet{Shih05} found a long-term modulation with a period of  30--40 days, which corresponds
to the regular transition between the hard and soft states \citep{Belloni07}.

The \textit{XMM-Newton} observations sample different parts of the diagrams shown in Figure~\ref{fig:cc}. During Obs.~1 and 6, the intensity was low and the spectrum was hard (see Table 4 and 5 in \citealt{Sanna13} for more details on the spectral properties). Observations 2 to 5 were all done when the source was bright, and the source spectrum was relatively soft. The shape of the colour-colour diagram in the top panel of Figure~\ref{fig:cc} shows that 4U 1636--53 belongs to the so-called Atoll class \citep{Hasinger89}. The upper right corner of the diagram represents times when the source is in the so-called transitional state. As the mass accretion rate increases, the source first moves down to the left of the diagram, and then to right to the soft state \citep[see, e.g.,][]{vdKlis06}. The \textit{XMM-Newton} observations did not uniformly sample the full range of source colours and intensities, but covered two small regions of the diagram. Obs.~1 and 6 sampled the source in the transitional state, while Obs.~2--5 covered the soft state. The position of the source on the colour-colour diagram is parameterised by the length of the coordinate $S_a$ \citep{Mendez99}, which approximates the shape of the diagram with a curve. The length of $S_a$ is arbitrarily normalised to the distance between $S_a=1$ at the top right corner, and $S_a=2.5$ at the bottom right corner, via $S_a=2$ at the bottom left vertex of the colour-colour diagram (see Figure~\ref{fig:cc}). Similar to the $S_z$ coordinate in the Z sources \citep{Vrtilek90}, $S_a$ is considered to map mass accretion rate \citep[see e.g.,][]{Hasinger89,Kuulkers94, Mendez99}. As reported in \citet{Sanna13}, the order of the six \textit{XMM-Newton/RXTE} observations, going from the transitional state to the soft state, according with their $S_a$ values is: 1--6--2--3--5--4.

\begin{table*}
\caption{Inner radius inferred from the iron line measurements sorted as a function of the $S_a$ parameter (from \citealt{Sanna13})}
\begin{tabular}{|c|c|c|c|c|c|c|}
     &  Obs.~1 &  Obs.~6 &  Obs.~2 & Obs.~3 & Obs.~5 & Obs.~4 \\
     &  $R_{in}$(GM/c$^2$) &  $R_{in}$(GM/c$^2$)& $R_{in}$(GM/c$^2$) & $R_{in}$(GM/c$^2$)&$R_{in}$(GM/c$^2$) & $R_{in}$(GM/c$^2$)\\     
\hline
\textsc{diskline}   & $10.6^{+1.5}_{-2.6}$ & $8.0^{+6.3}_{-2.0*}$    & $10.7^{+4.5}_{-2.4}$ & $8.4^{+0.7}_{-1.5}$ & $6.5^{+0.6}_{-0.5*}$ &  $6.0^{+2.8}_{-0.0*}$ \\\\
\textsc{laor}       &$10.8^{+0.6}_{-2.9}$   & $6.2\pm1.9$    & $4.0^{+5.6}_{-0.8}$ &  $2.3^{+0.2}_{-0.5}$ & $2.0\pm{0.3}$&$2.8^{+1.2}_{-0.8}$ \\\\
\textsc{kyrline} a$_*$=0  & $10.8^{+2.0}_{-1.3}$    & $12.2^{+1.9}_{-2.6}$  &$6.3^{+1.1}_{-0.3*}$  & $13.1\pm{1.3}$ &$6.2^{+0.4}_{-0.2*}$& $6.0^{+1.9}_{-0.0*}$  \\\\
\textsc{kyrline} a$_*$=0.27  & $10.6^{+1.9}_{-1.2}$    &$12.2^{+1.7}_{-2.6}$   & $12.5^{+2.9}_{-2.0}$  & $13.1\pm{1.4}$ & $5.9^{+0.4}_{-0.8*}$ & $5.7^{+2.0}_{-0.6*}$  \\\\
\textsc{kyrline} a$_*$=1  & $9.9^{+1.9}_{-1.1}$  &$12.1\pm{2.1}$  & $11.8^{+3.2}_{-1.7}$  & $12.9^{+3.2}_{-1.7}$ &  $2.6\pm0.1$&$5.3^{+1.4}_{-1.9}$ \\\\
\textsc{reflection}    & $12.6\pm{1.7}$  &$19.1^{+7.6}_{-10.8}$  & $7.8^{+3.1}_{-2.7}$  & $15.4\pm2.7$ &  $5.4\pm0.1$&$5.6^{+2.2}_{-0.3}$  \\
\hline
\end{tabular} \\
\flushleft Notes: A * means that the radius pegged at the hard limit of the range.
\label{tab:line_radius}
\end{table*}


\subsection{Iron line and measurements of the inner accretion radius}

In Table~\ref{tab:line_radius} we report the values
of the inner disc radius for the different models used by \citet{Sanna13} to fit the line. The observations in Table~\ref{tab:line_radius}
are sorted according to their $S_a$ values. As already noted in \citet{Sanna13}, the inner disc radius
inferred from the relativistic profile of the iron line in 4U 1636--53 does not change in correlation with the position in 
the colour-colour diagram contrary to what is predicted by the standard accretion disc model \citep[see e.g.,][and references therein]{Done07}. 
\citet{Sanna13} also found that other parameters used to fit the line, such as line energy, source inclination and equivalent 
width, do not show any clear correlation with the source state \citep[see Figure~6 in][]{Sanna13}.

\subsection{kHz QPOs and measurements of the inner accretion radius}
\label{sec:qpo_analysis}
\begin{figure*}
\begin{center}$
\begin{array}{ccc}
\includegraphics[scale=0.41]{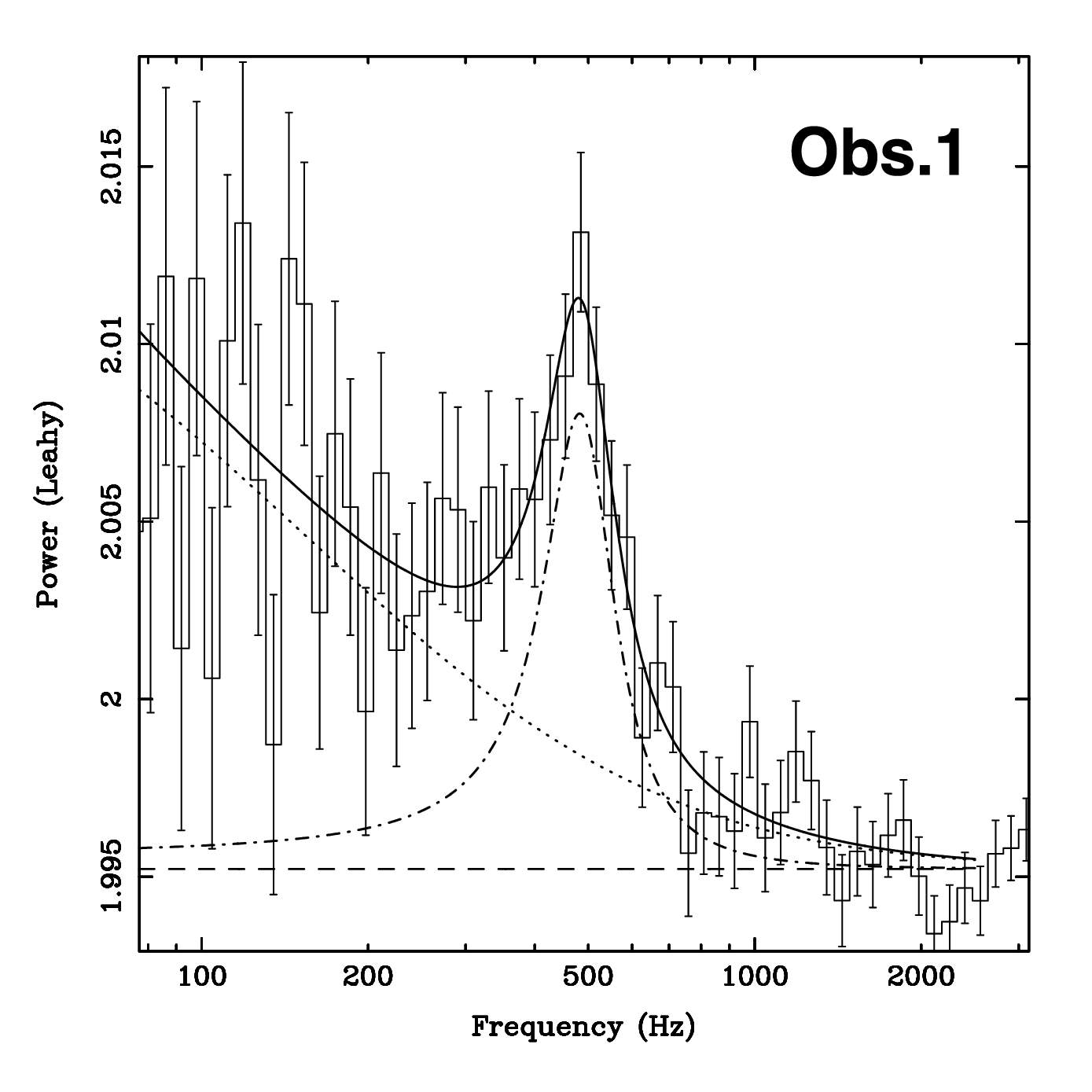} & 
\includegraphics[scale=0.41]{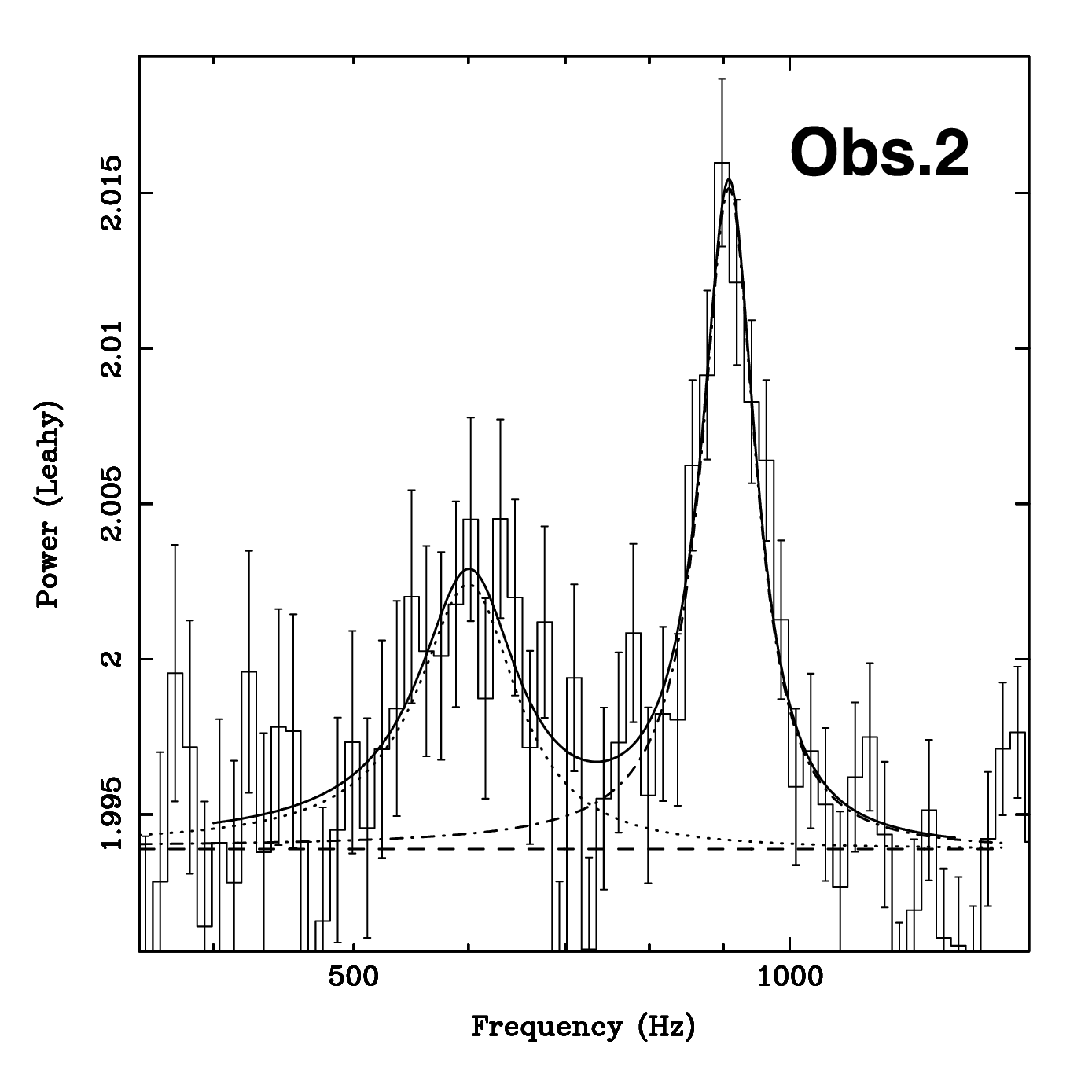}&
\includegraphics[scale=0.41]{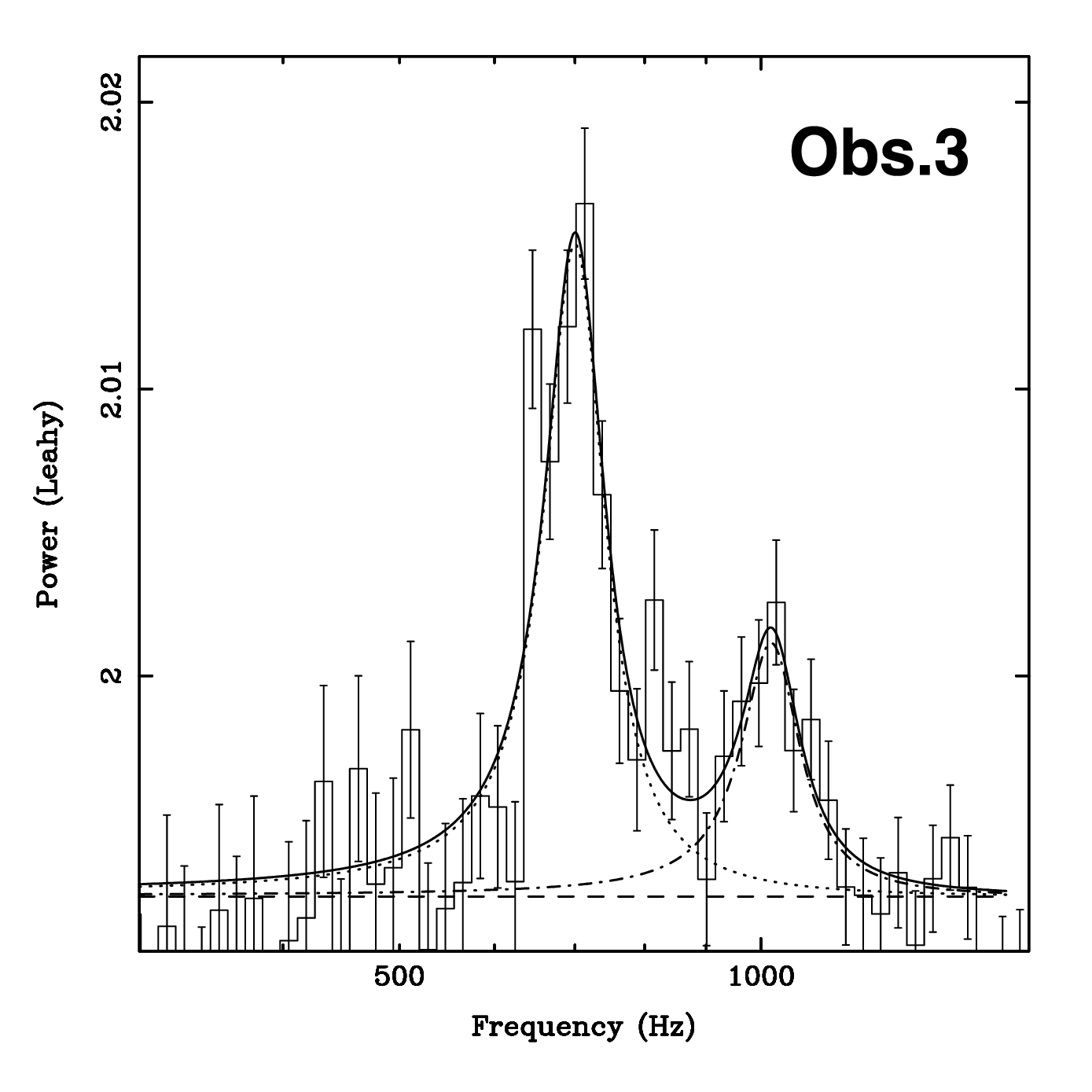}\\
\includegraphics[scale=0.41]{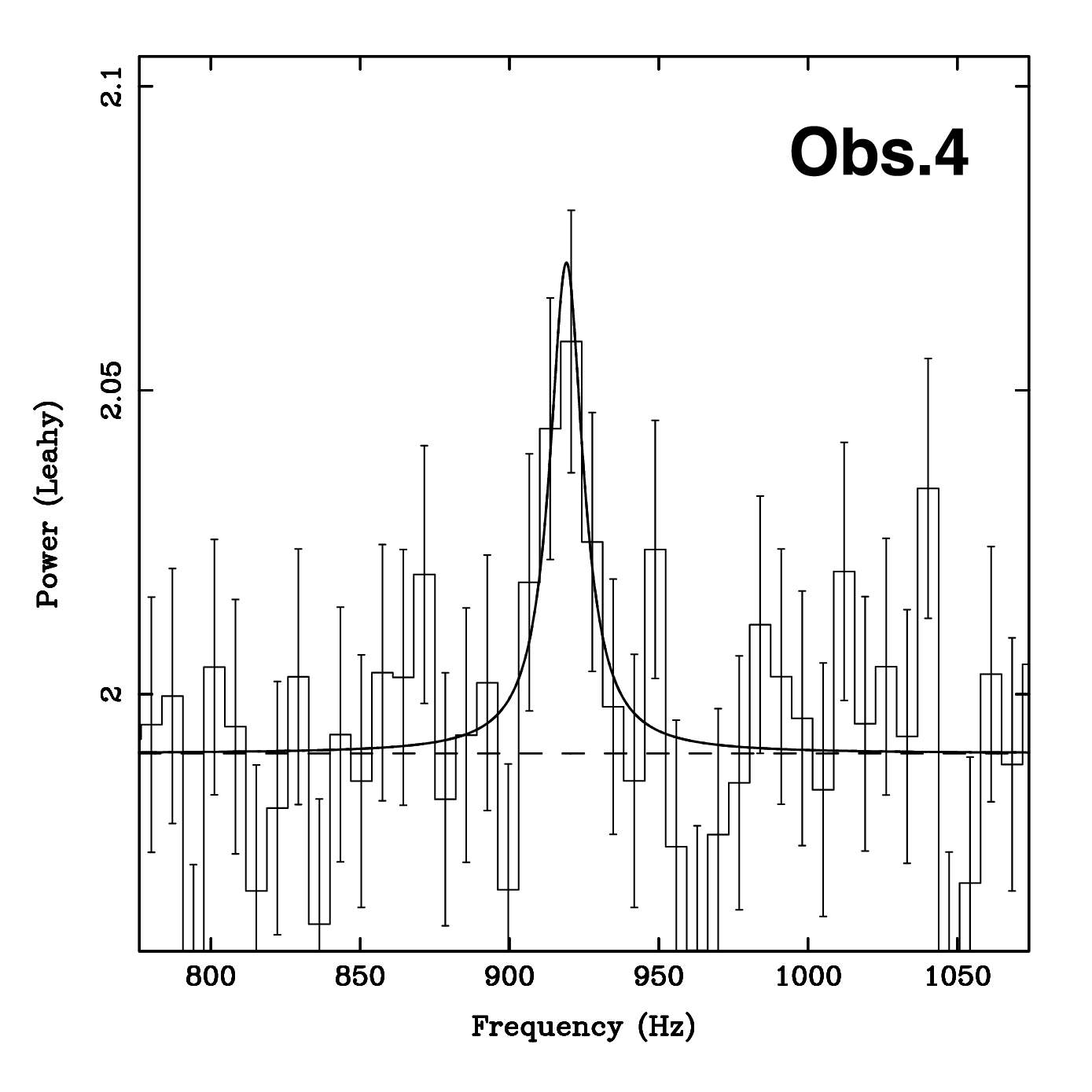} & 
\includegraphics[scale=0.41]{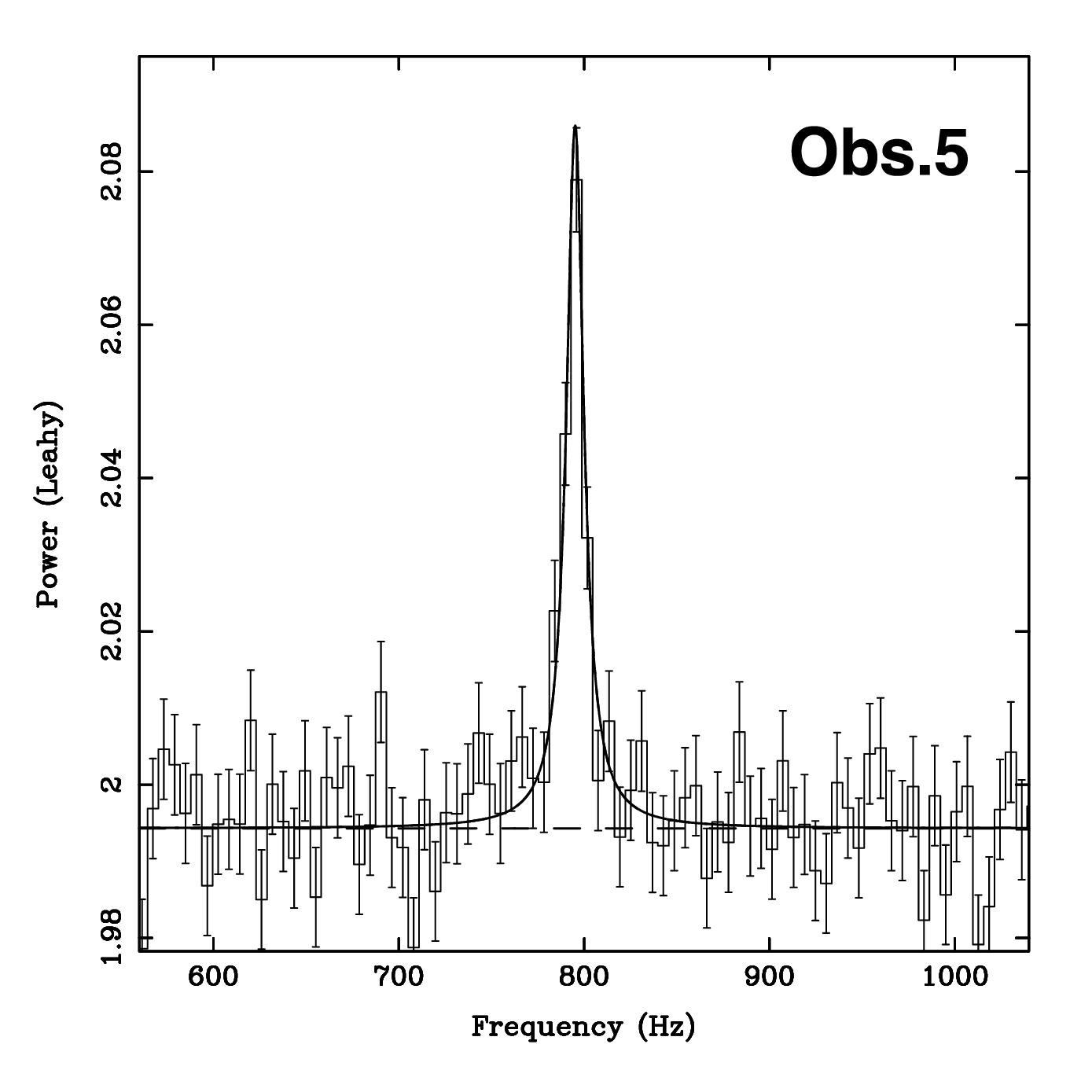}&
\includegraphics[scale=0.41]{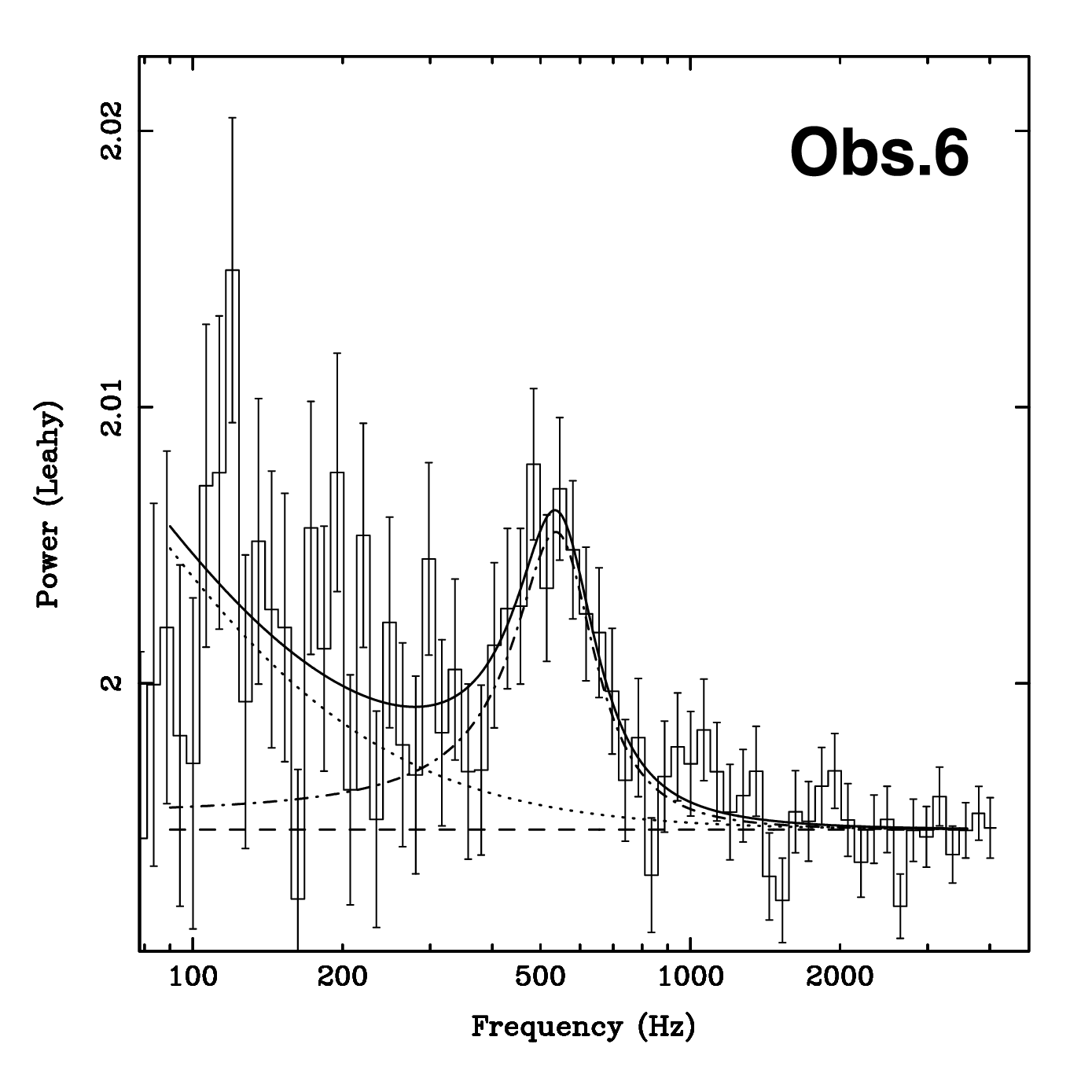}\\
\end{array}$
\end{center}
\caption{Leahy normalised power density spectra for Obs.~1--6 of 4U 1636--53, calculated from the \textit{RXTE} observations.
The power density spectra were fitted with a model consisting of a constant, one or two Lorentzians to fit the kHz QPOs, and
(if required) a Lorentzian to model the residual broad band noise at low frequencies. For Obs.~5 we show the kHz QPO after we applied the shift-and-add method \citep[see][for more details]{Mendez98}. }
\label{fig:qpo}
\end{figure*}

We detected one or two kHz QPOs in the average PDS of each \textit{RXTE} observation that 
was performed simultaneously with an \textit{XMM-Newton} observation. In Obs.~1 and 6 we detected a
strong broad-band noise component extending up to few hundred Hz plus a single kHz QPO at 
$\sim$480 and $\sim$540 Hz, respectively. The kHz QPOs in Obs.~1 and 6 have rms 
amplitudes of $\sim$12\% and $\sim$14\%, respectively. The overall power-spectral
shape in both cases (not shown) is similar to those previously observed in the transitional state of 4U 1636--53
\citep[e.g., intervals A--C in][]{Altamirano08a} and other sources \citep[e.g.,][]{Straaten02,Straaten03}.
In Obs.~4 we detected a single kHz QPO at $\sim$920 Hz, with an rms amplitude of about 7\%. 
Twin kHz QPOs at $\sim$600 and $\sim$905 Hz are present in Obs.~2 with rms amplitudes of $\sim$7\% and $\sim$10\%,
respectively, and at around 700 and $\sim$1020 Hz in Obs.~3, both with rms amplitudes of $\sim$6\%. In Obs.~5 we
detected two peaks with a frequency separation significantly lower than the average frequency difference 
between the lower and the upper kHz QPOs previously reported for this source \citep{Mendez98,Jonker02,Mendez07,Altamirano08a}. We investigated the evolution of the QPO frequency with time in this observation, and we found that the QPO
signal appears sporadically during the long observation, and when it is detectable the frequency changes with 
time between $\sim$750 Hz and $880$ Hz. It is likely that Obs.~5 showed only one kHz QPO that
moved in frequency, we therefore applied the shift-and-add
method \citep{Mendez98} to recover the properties of the QPO. Note that we did not find the upper kHz QPO (see Figure~\ref{fig:qpo}), and that
 the frequency reported for the kHz QPO in Obs.~5 is the weighed average of the frequencies spanned by the QPO during the observation. 
In Figure~\ref{fig:qpo} we show the Leahy normalised power density spectra at high frequencies for the six observations, and in Table~\ref{tab:kHz}
we give the best-fitting parameters of the kHz QPOs.

\begin{figure}
\resizebox{1\columnwidth}{!}{\rotatebox{0}{\includegraphics[clip]{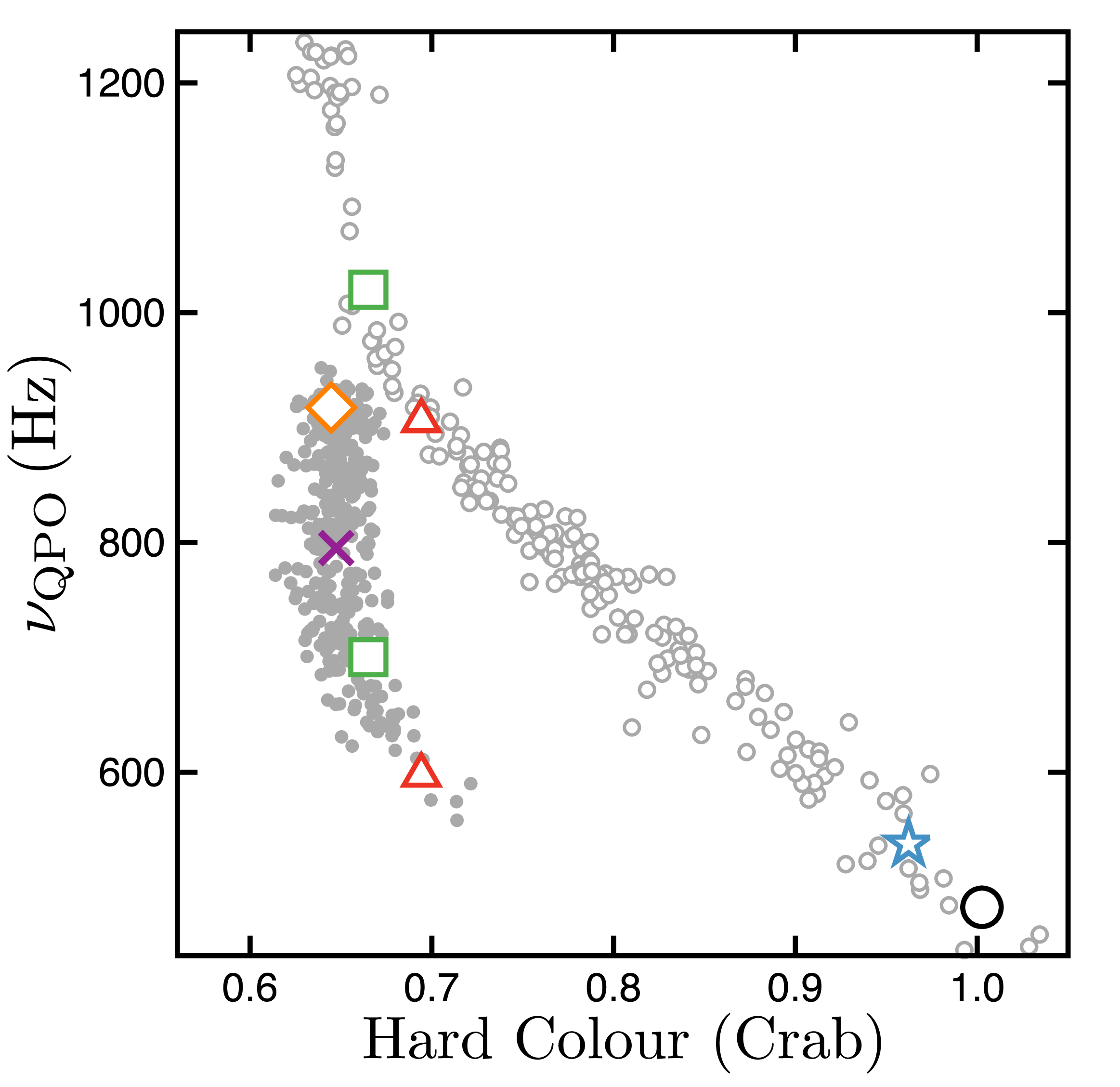}}}
\caption{Frequency of the kHz QPOs in 4U 1636--53 as a function of hard colour. Grey filled
and empty bullets represent the lower and the upper kHz QPOs, respectively \citep[see][]{Sanna12a}. With large symbols we
represent the kHz QPOs detected in Obs.~1--6. The symbols have the same meaning as in Figure~\ref{fig:cc}.}
\label{fig:freq-hard}
\end{figure}

In those observations in which we detected two simultaneous QPOs we can readily identify the lower and the upper kHz QPOs.
For the other observations we used the frequency vs. hard colour digram that, as shown in \citet[][see also \citealt{Belloni07}]{Sanna12a}, can be
used to identify the lower and upper kHz QPOs in 4U 1636--53. 
In Figure~\ref{fig:freq-hard} we show the centroid frequency of the kHz QPOs detected in 4U 1636--53 as a function
of hard colour, with different symbols for the lower (grey filled bullets) and the upper (grey empty bullets) kHz QPOs. 
On top of that we show the kHz QPOs detected in Obs.~1--6.  As expected, Figure~\ref{fig:freq-hard} confirms
that the two simultaneous QPOs detected in Obs.~2 and 3, are indeed the lower and the upper kHz QPOs. 
Figure~\ref{fig:freq-hard} also shows that the QPOs detected in Obs.~1 and 6 are both upper kHz QPOs,
while the QPOs in Obs.~4 and 5 are both the lower kHz QPOs.

\begin{table*}
\caption{Parameters of the kHz QPOs in 4U 1636--53 sorted as a function of the $S_a$ parameter}
\begin{tabular}{|c|c|c|c|c|c|c|c|}
         &     &  Obs.~1 &  Obs.~6 &  Obs.~2 & Obs.~3 & Obs.~5 & Obs.~4 \\
\hline
\multirow{3}{*}{$L_{\ell}$} & $\nu$ (Hz)  & -- &--& $597\pm13$    & $700\pm5$ & $795.1\pm0.5$ & $917\pm3$   \\
                           & FWHM (Hz)  & -- & -- & $115^{+42}_{-29}$    & $99\pm14$ & $11.5\pm1.3$  &$15\pm6$  \\
                           & rms (\%)   & -- & -- & $6.7\pm0.8$  & $8.5\pm0.4$ &  $11.7\pm0.3$ &  $6.8\pm0.9$          \\
\hline
\multirow{3}{*}{$L_{u}$} & $\nu$ (Hz) &$482\pm11$  &$537\pm18$ &  $906\pm5$   & $1020\pm14$      & -- & -- \\
                           & FWHM (Hz) & $178\pm45$& $247^{+72}_{-58}$ & $99\pm13$   & $140^{+48}_{-37}$ &  -- & -- \\
                           & rms (\%) & $11.5\pm1.5$&$14.1\pm1.4$  & $9.8\pm0.5$ & $6.3\pm0.7$    & -- & -- \\
\hline
\end{tabular} \\
\flushleft Notes: $L_{\ell}$ and $L_{u}$ stand for the lower and the upper kHz QPO, respectively. For Obs.~5 we report the kHz QPO parameters after applying the shift-and-add method \citep{Mendez98}.
\label{tab:kHz}
\end{table*}


In order to investigate whether the frequency of the detected kHz QPOs remains approximately constant in time, 
we studied the dynamical power spectra \citep[e.g.,][]{Berger96} using different frequency and time binning factors. 
We were unable to follow the time evolution of the upper kHz QPO in any of the observations. We were able to trace the frequency of the lower kHz QPO only in two cases: During Obs.~3 the QPO frequency varied between $\sim$620 Hz and
$\sim$810 Hz while, as already mentioned, during Obs.~5 the QPO moved between $\sim$750 Hz and $\sim$880 Hz. 

In accordance with the scenario of an accretion disc truncated at a larger radii in the
hard than in the soft state, and a disc extending very close to the NS surface in the soft state \citep{GierlinskiDone02, Done07},
the frequency of the upper kHz QPO is lower (larger inner radius) in Obs.~6 and 1 (hard state) than in Obs.~2 and 3 (soft state).

\subsection{Iron lines and kHz QPOs as tracers of the inner radius of the accretion disc}

Both kHz QPOs and relativistically-broadened iron lines likely reflect properties of
the accretion flow in the inner edge of the accretion disc \citep[e.g.,][]{Fabian89,MillerLambPsaltis98}. To investigate whether the two interpretations match,
we compared the inner radius estimated from the frequency of the kHz QPO and the profile of the iron line
when both were detected simultaneously. 
Most models predict that the upper kHz QPO frequency in LMXBs represents the orbital 
frequency at the inner edge of the accretion disc \citep[e.g.,][]{MillerLambPsaltis98,Stella98}. The expression 
for the orbital frequency in the space time outside a slowly and uniformly rotating NS is 
$\nu_{\phi}=\nu_k(1+a_*(R_g/r)^{3/2})^{-1}$ where $\nu_k = (1/2\pi)\sqrt{GM/r^3}$ is the Keplerian frequency,
$a_*=Jc/GM^2$ is the angular momentum parameter, $R_g=GM/c^2$ is the gravitational radius,
 \textit{G} is the gravitational constant, \textit{M} the mass of the NS star, \textit{r} the radial 
distance from the center of the NS star, and \textit{c} is the speed of light.

In order to estimate the inner radius we need both the angular momentum parameter and the NS mass. As
reported in \citet{Sanna13}, taking into account the spin frequency of 581 Hz of the NS in 4U 1636--53 \citep{Zhang97,Giles02,Strohmayer02},
we estimated  $a_*$ to be $~0.27$. On the other hand, the mass of the NS in this system is unknown, therefore
 we calculated the inner radius of the accretion disc, $R_{in}$, as a function of the NS mass. The lack of information on the NS
mass does not allow us to directly compare $R_{in}$ inferred from the kHz QPO and the iron line simultaneously, 
however we can compare the trend of $R_{in}$ as a function of other properties of the source, e.g., position in the 
colour-colour diagram or intensity, and test whether there is a single value of the mass of the NS for which the two
different estimates of the inner disc radius agree.

\begin{figure}
        \centering
         \caption{}
        \begin{subfigure}[b]{1\columnwidth}
                \centering
                \includegraphics[width=0.75\textwidth]{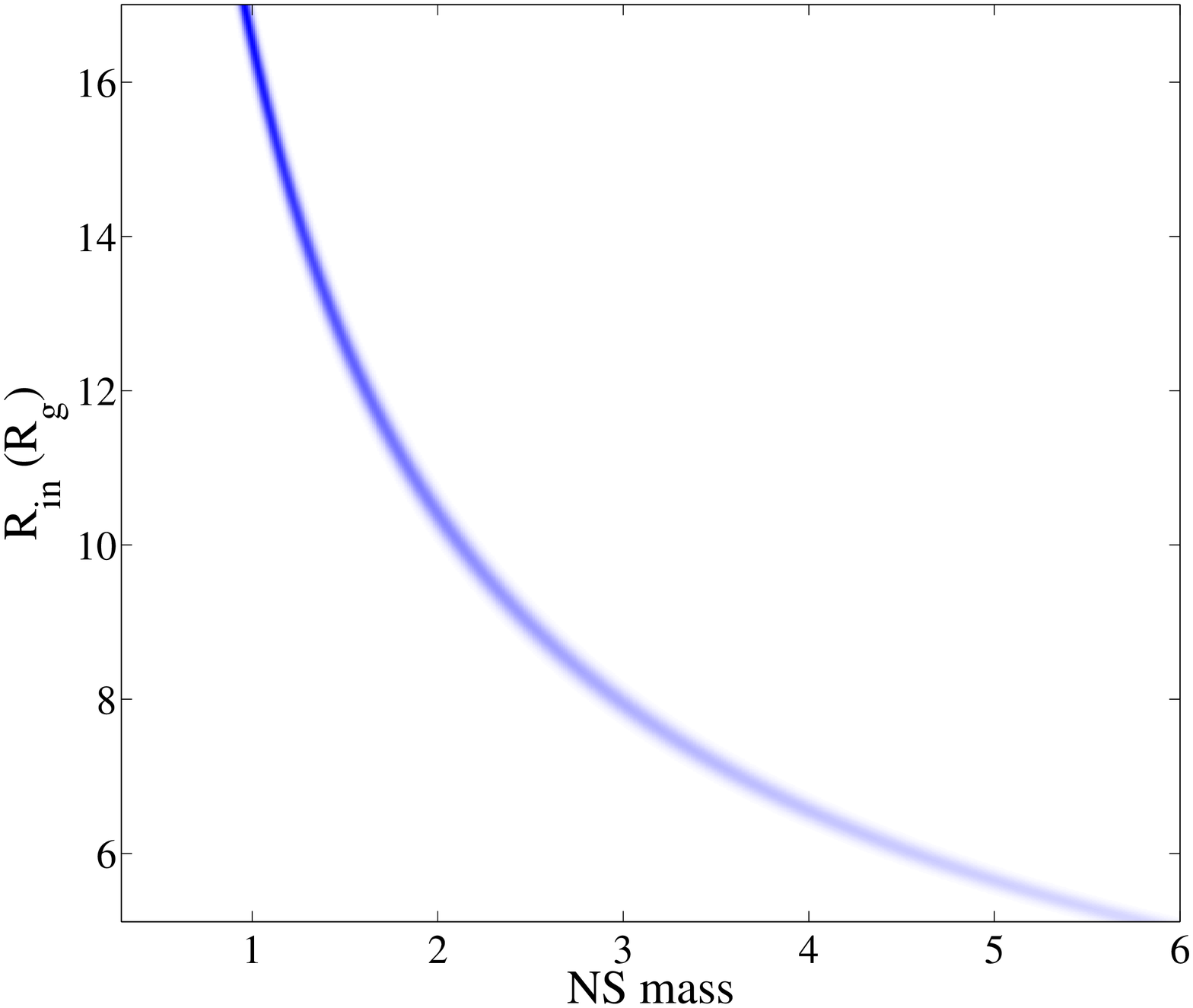}
                \caption{Probability distribution function of the inner radius of the accretion disc as a function NS mass in 4U 1636--53, inferred from the upper kHz QPO in Obs.~1. Colour density represents the confidence level.}
                \label{fig:radius_qpo}
        \end{subfigure}\\
        \begin{subfigure}[b]{1\columnwidth}
                \centering
                \includegraphics[width=0.75\textwidth]{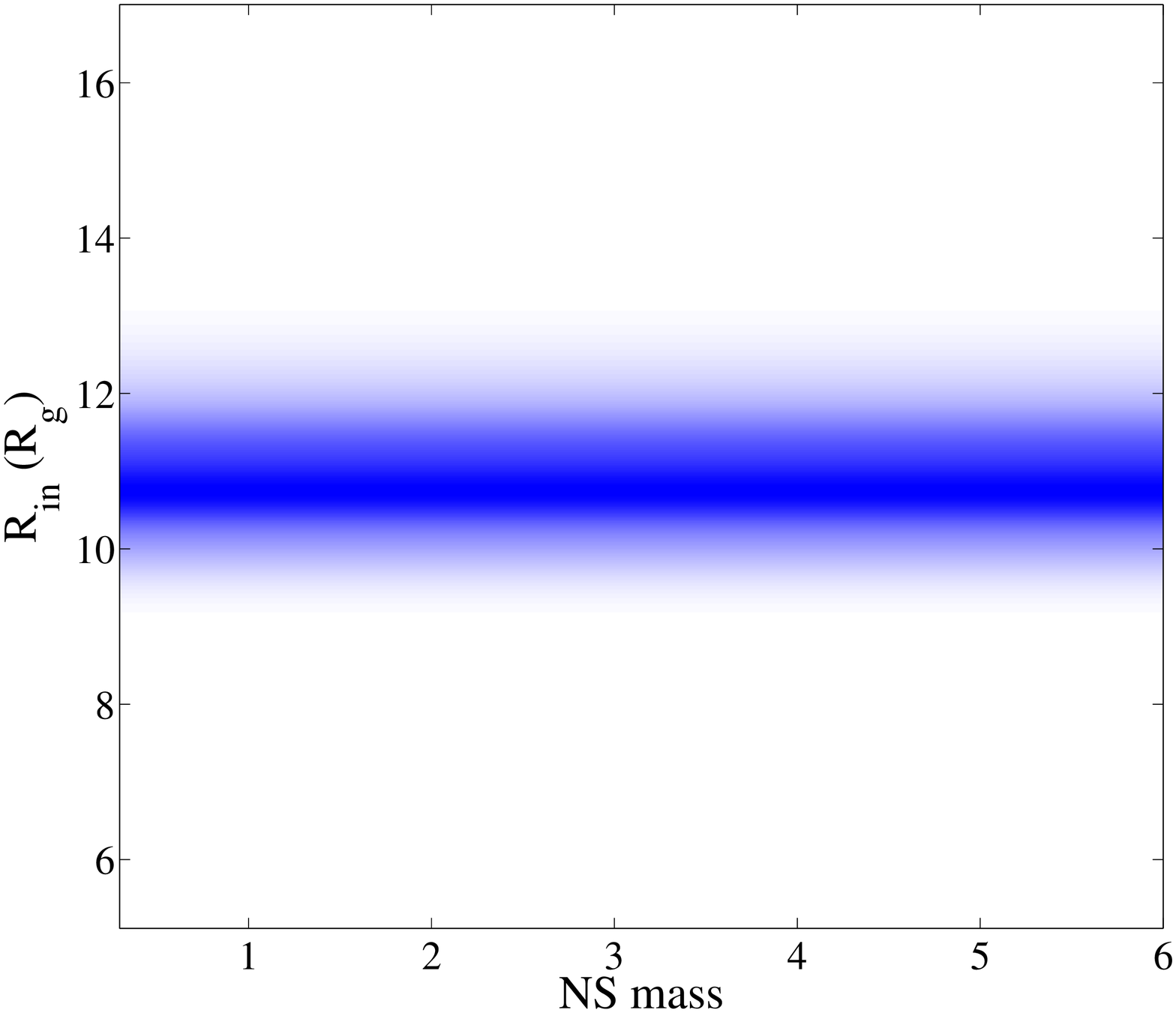}
                \caption{Probability distribution function of the inner radius of the accretion disc as a function of NS mass in 4U 1636--53, inferred from the fit of the iron emission line in Obs.~1 using the relativistic line model \textsc{kyrline} with $a_*=0.27$. Colour density represents the confidence level. }
                \label{fig:radius_line}
        \end{subfigure}\\
        ~ 
        \begin{subfigure}[b]{1\columnwidth}
                \centering
                \includegraphics[width=\textwidth]{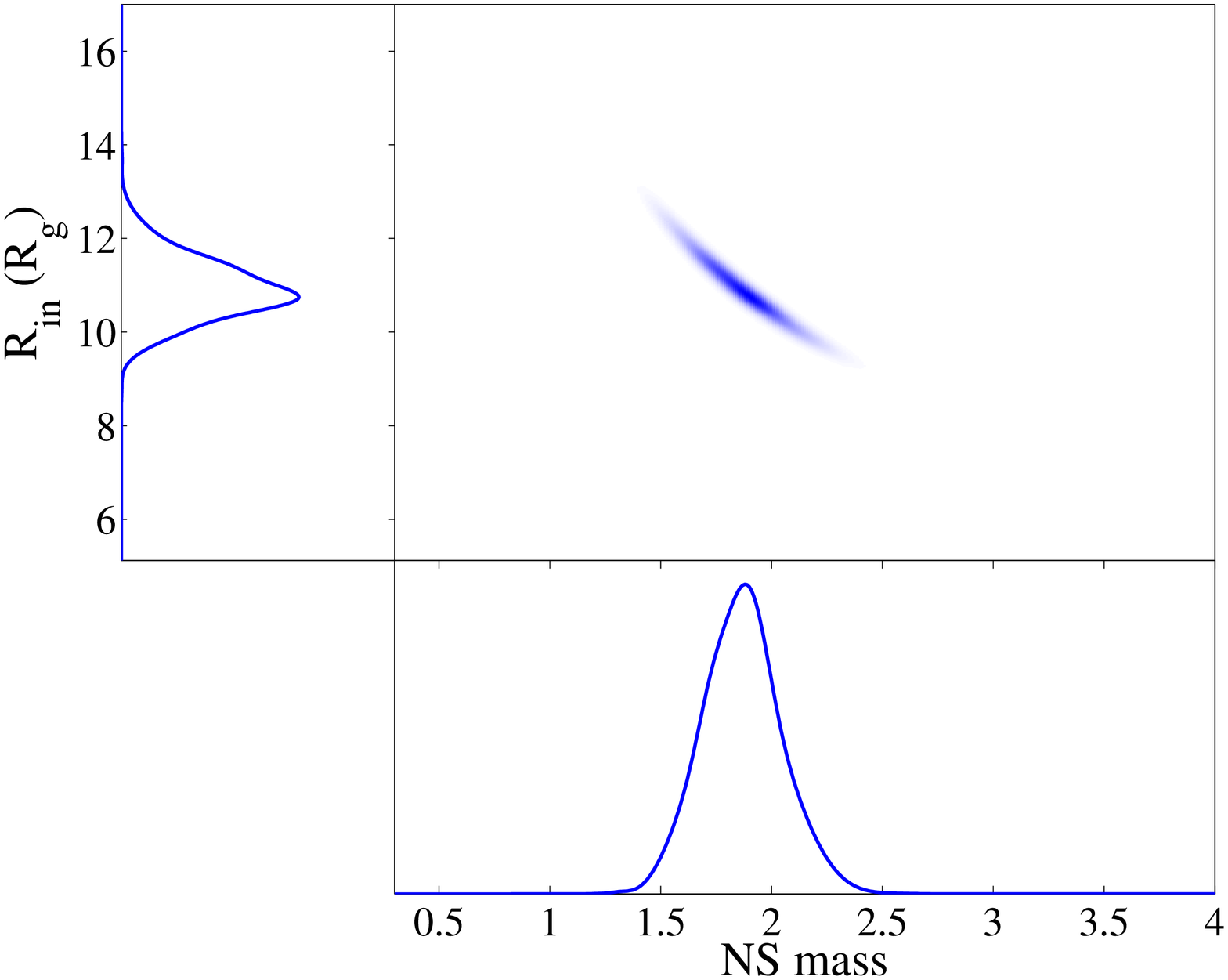}
                \caption{Joint probability distribution function (top-right panel), marginal probability distribution functions for the inner disc radius (top-left panel) and the NS mass (bottom panel) for 4U 1636--53, calculated combining the probability density distribution functions from Figures~\ref{fig:radius_qpo} and \ref{fig:radius_line}. }
                \label{fig:joint_prop}
        \end{subfigure}       \label{fig:qpo_lines}
\end{figure}

To explain in more detail how we compared the simultaneous timing and spectral information, in Figure~\ref{fig:qpo_lines}
we show a step-by-step example for the case of Obs.~1. As reported in Table~\ref{tab:kHz}, Obs.~1 showed 
an upper kHz QPO at a frequency of $\sim$480 Hz. In Figure~\ref{fig:radius_qpo} we show the inner radius of the accretion disc
 in units of $R_g$ versus the NS mass inferred from the relation between $\nu_{\phi}$, $M$ and $R_{in}$. 
Colour density represents the probability density distribution (PDF) of the 
inner radius taking into account the uncertainties from the fit of the QPO frequency. In Figure~\ref{fig:radius_line} we show
the inner radius in units of $R_g$ inferred from the iron emission line when we fitted it with the relativistic line model \textsc{kyrline} with $a_*=0.27$ \citep{Sanna13}.
Similar to Figure~\ref{fig:radius_qpo}, the PDF of the inner radius is shown in colour density.
Note that the radius derived from the iron line does not depend upon the NS mass; however, for practical purposes, we 
plotted the PDF of the inner radius using a similar layout as for the kHz QPO. 
In Figure~\ref{fig:joint_prop} (central panel) we show the joint probability distribution function of the inner radius and the NS mass derived from 
the iron line profile and the frequency of the kHz QPO. In Figure~\ref{fig:joint_prop} we also show the marginal distributions for the
NS mass and the inner disc radius in units of $R_g$, calculated by integrating the joint PDF
over the radius and the mass, respectively. 

From this example, we find that for this observation of 4U 1636--53 the kHz QPO frequency
and the iron line modelling are consistent for a NS mass of $\sim$1.9 M$_\odot$, and an 
accretion disc extending down to $\sim$11 R$_g$; the NS mass is slightly high,
but still consistent with most NS equations of state \citep[see, e.g.,][]{Lattimer07}. The fact that the NS mass estimate is consistent 
with theoretical expectations suggests that, for this particular case, the kHz QPO and iron line interpretations both hold.
 Since we have observations mapping different positions in the colour-colour diagram, we can test whether the kHz QPOs and
 the Fe line give a consistent value of the NS mass for all accretion states. To do that we used the four observations,
 two in the transitional state (Obs.~1 and 6) and two in the soft state (Obs.~2 and 3), where
we detected both the upper kHz QPOs and the broad iron line simultaneously. For completeness, we used the inner radius
estimated from fits of the iron line profile with different line models \citep[see Table~\ref{tab:line_radius} for the list of models; see][for a discussion
of the line models]{Sanna13}. 

In Figure~\ref{fig:masses} we show, for different line models, the marginal probability distribution functions of the NS mass for the four
observations with both the upper kHz QPOs and the iron emission line. The mass values derived
from different observations do not yield consistent results, regardless of the iron line model. We further notice
that the NS mass values in 4U 1636--53 derived from this method span a range between $\sim$0.5 M$_\odot$ and $\sim$3.0 M$_\odot$,
with the exception of the mass inferred from the fits of the line with the \textsc{laor} model, which spans a wider range (1.2--10 M$_\odot$). 

\section{discussion}

We detected kHz QPOs in all the \textit{RXTE} observations simultaneous with the six \textit{XMM-Newton} observations of the NS LMXB 4U 1636--53 for which \citet{Sanna13} studied the broad iron line in the X-ray spectrum. Combing the measurements of the frequency of the kHz QPOs and the parameters of the iron lines in 4U 1636--53 we investigated the hypothesis that both the iron line and the kHz QPOs originate at (or very close to) the inner radius of the accretion disc in this system. 
 From these observations we found that the inner disc radius, deduced from the upper kHz QPO frequency,
 decreases as the spectrum of the source softens, and the inferred mass accretion rate increases. 
 On the other hand, the inner radius estimated from the modelling of the relativistically-broadened iron line 
 did not show any clear correlation with the source state, except for the line model \textsc{laor} for which the inferred 
 inner disc radius consistently decreases going from the transitional state to the soft state (see Table~\ref{tab:line_radius}). 
 Combining the disc radius inferred from the frequency of the upper kHz QPO and the iron line profile, 
 we found that the mass of the NS in 4U 1636--53 deduced from the four observations are inconsistent with being the same. A similar conclusion was drawn by \citet{Cackett10} from the first three observations in the sample that we studied here.
 The latter result implies that either the upper kHz QPO frequency does not reflect the orbital frequency at the inner edge of the disc, the Fe line profile is not (only) shaped by relativistic effects, the models used to fit the iron line are incorrect, or the the kHz QPOs and the Fe line are not produced in the same region of the accretion disc. 

\begin{figure*}
\begin{center}$
\begin{array}{cc}
\includegraphics[scale=0.23]{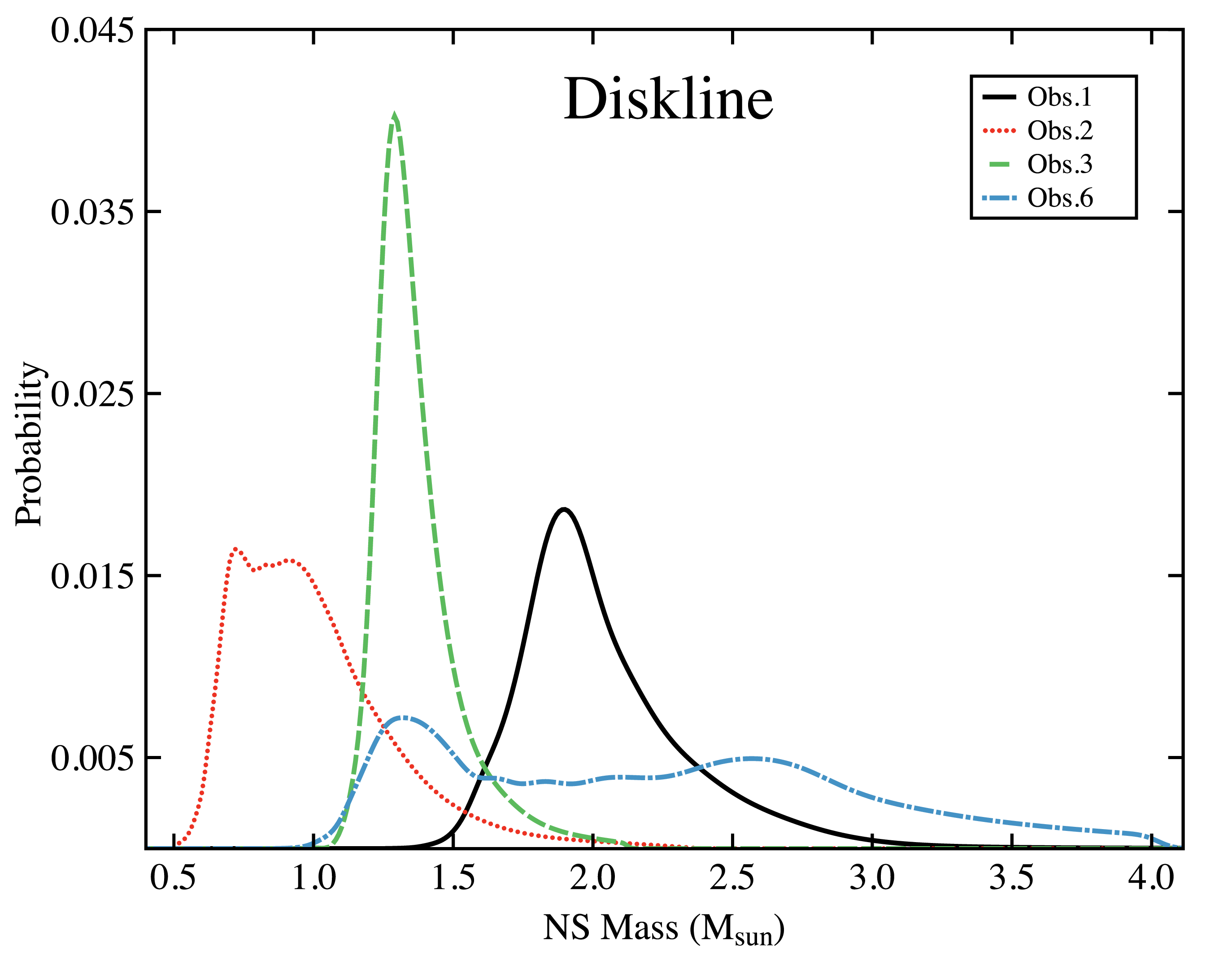} & 
\includegraphics[scale=0.23]{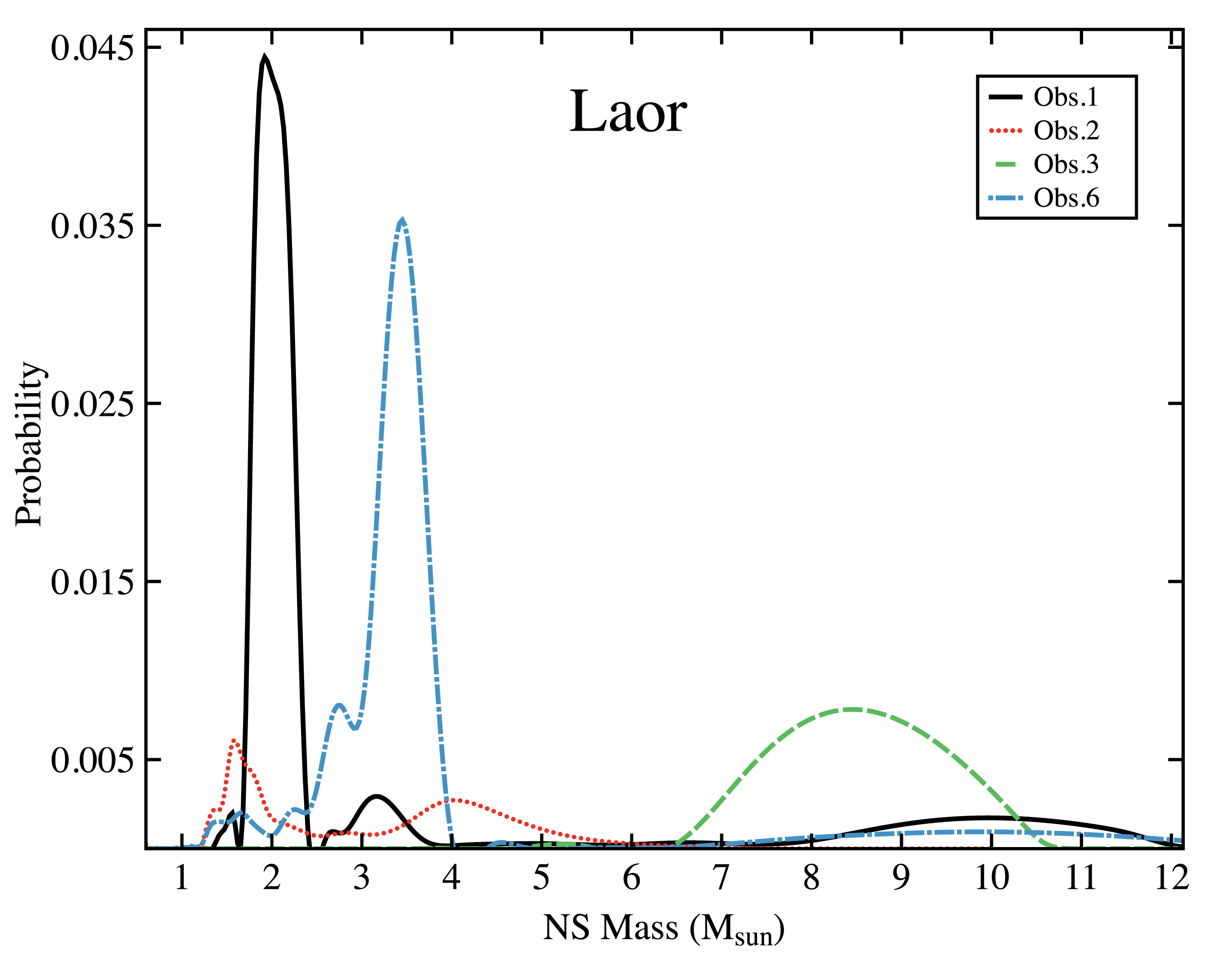}\\
\includegraphics[scale=0.23]{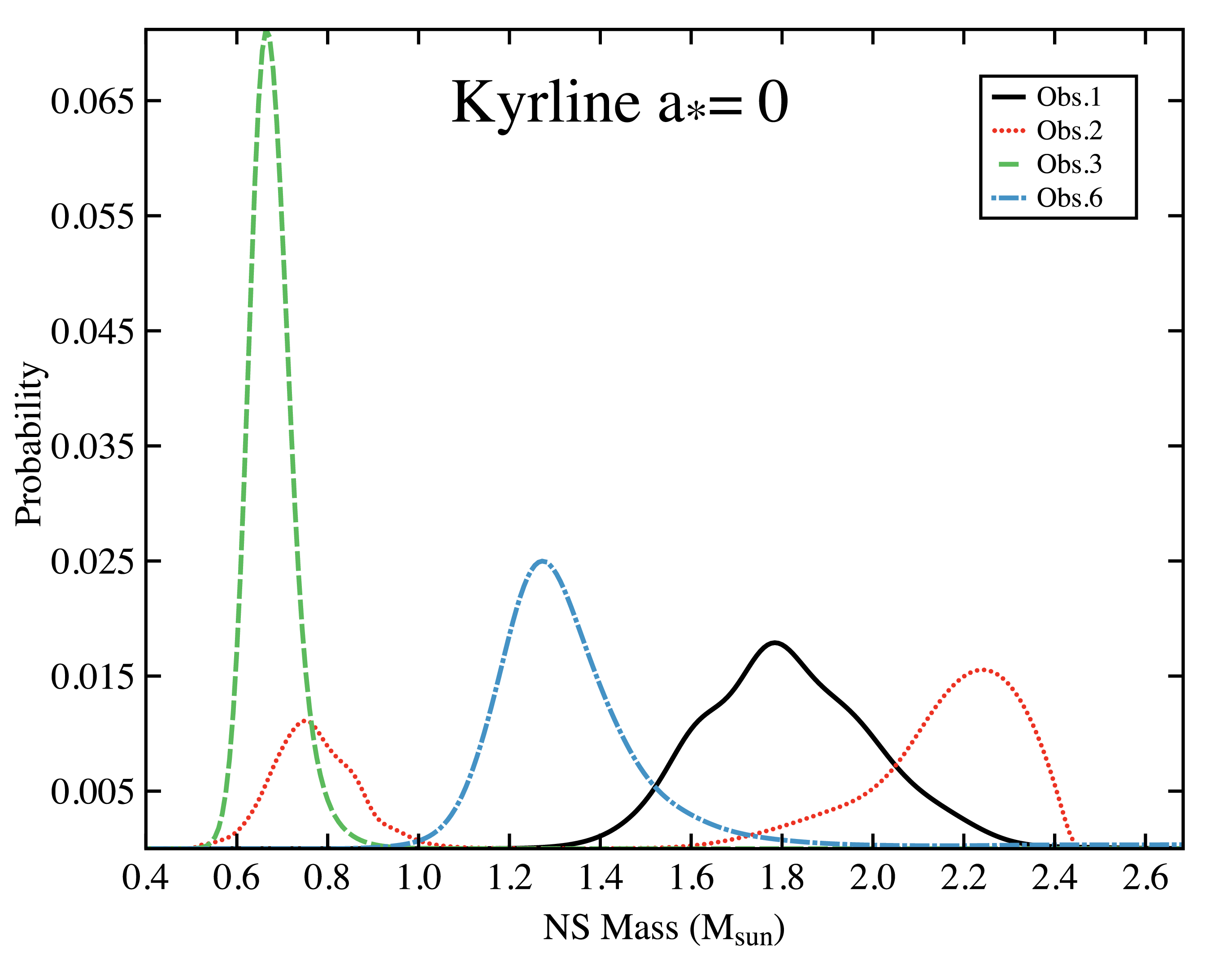}&
\includegraphics[scale=0.23]{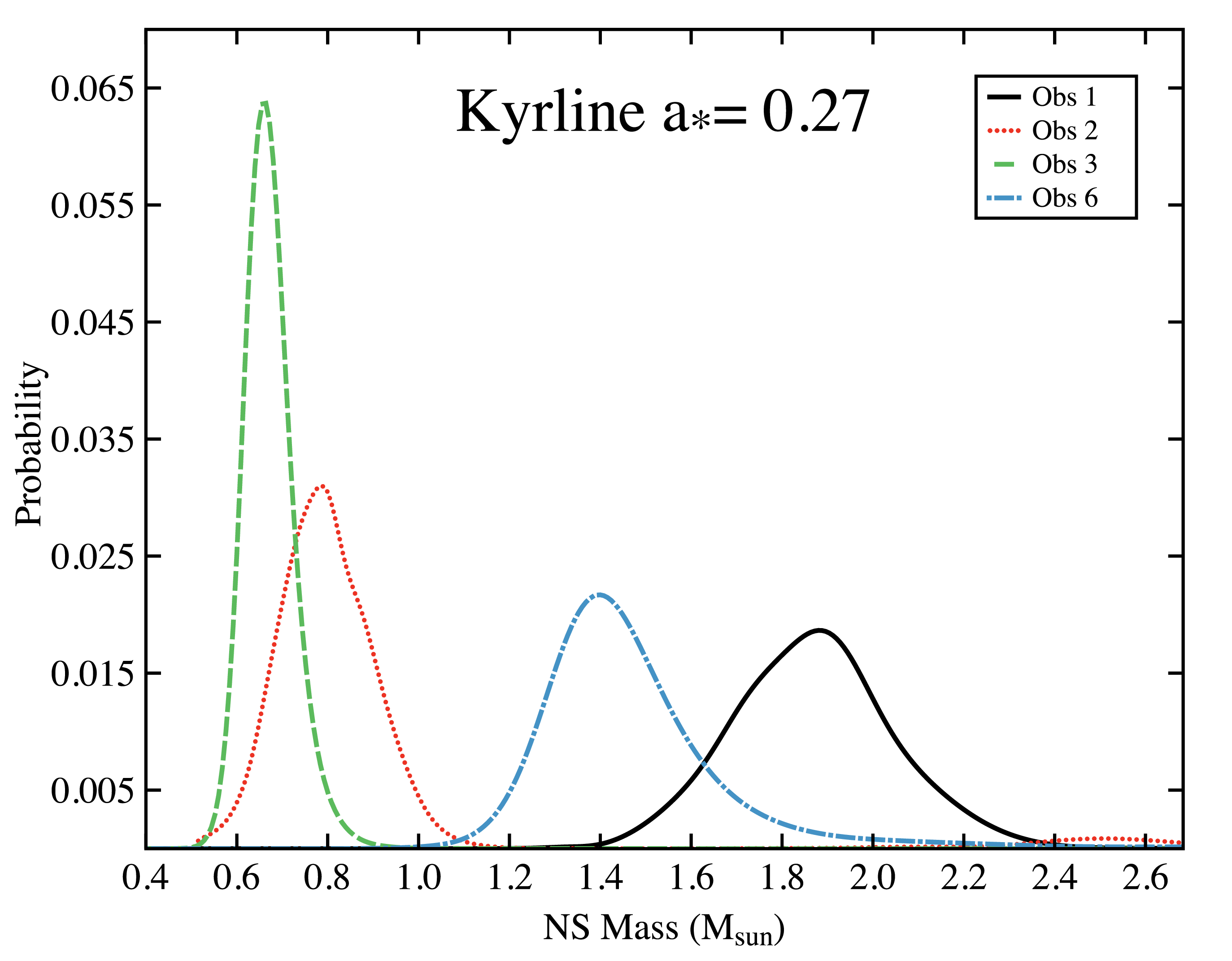} \\ 
\includegraphics[scale=0.23]{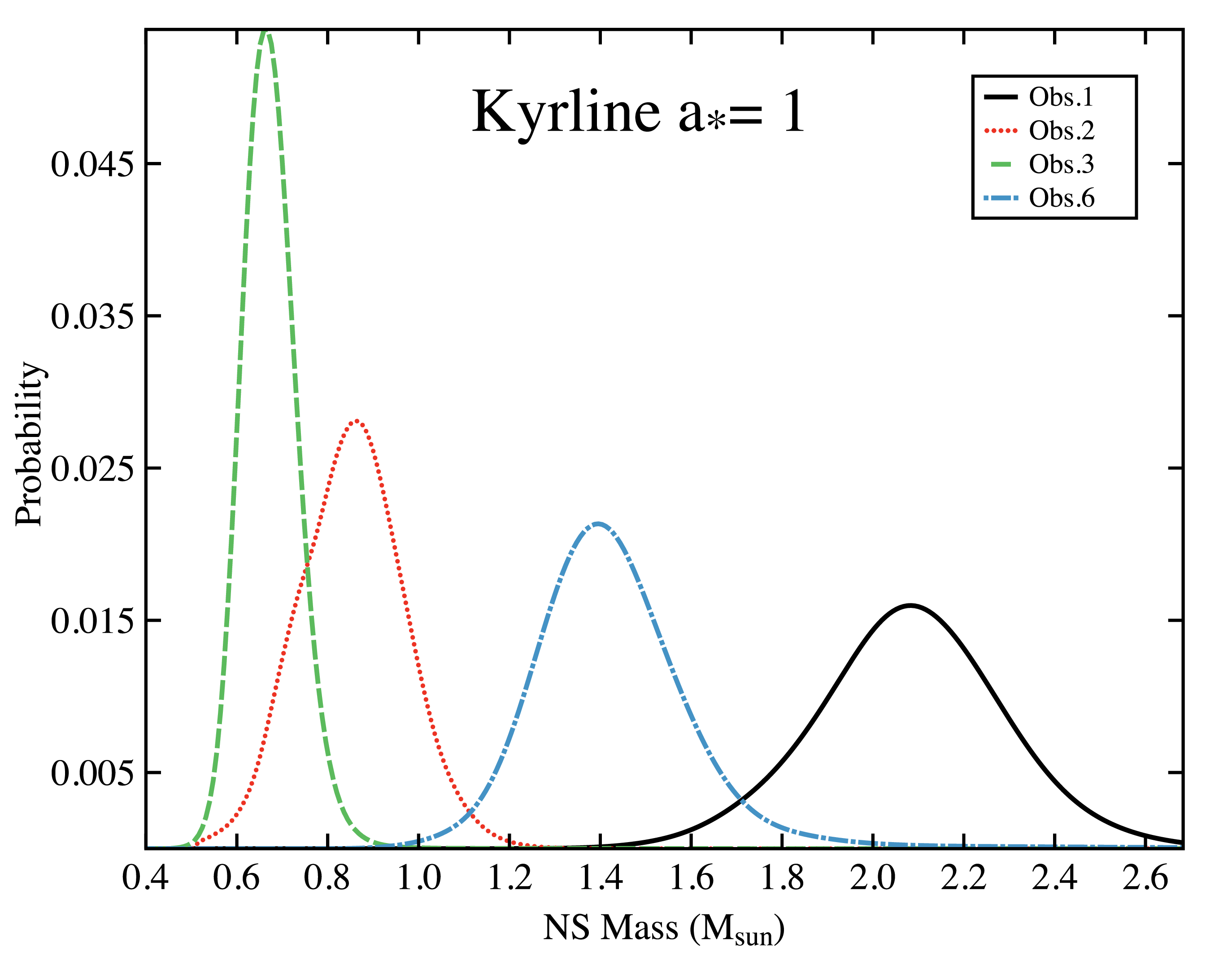}&
\includegraphics[scale=0.23]{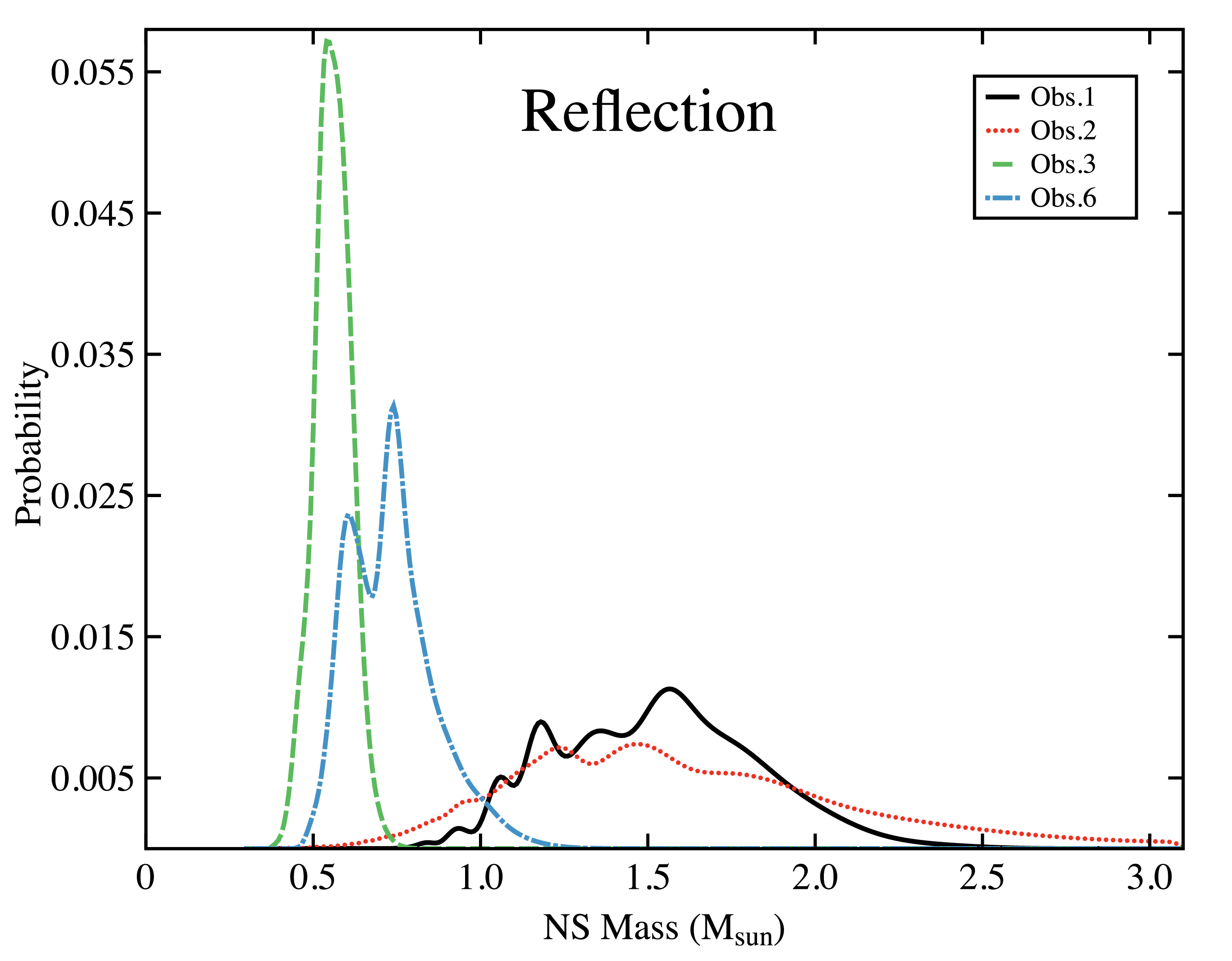}\\
\end{array}$
\end{center}
\caption{Marginal probability distribution functions of the NS mass in 4U 1636--53 inferred from simultaneous 
measurements of the upper kHz QPO and the iron emission line, for four different observations represented with different colours. Different panels represent
different models used to fit the iron line profile. }
\label{fig:masses}
\end{figure*}

We assumed that the upper kHz QPO is the one which reflects the orbital frequency at the inner edge of
the accretion disc \citep[e.g.,][]{MillerLambPsaltis98, Stella98}. There are, however, alternative models that associate
instead the lower kHz QPO to the orbital disc frequency \citep[e.g.,][]{Meheut09}. If this is the case, then the radius profile showed in
Figure~\ref{fig:radius_qpo} would shift to higher values of $R_{in}$, and therefore, the mass for which $R_{in}$ from kHz QPOs and 
iron lines would match will also shift to higher values. Since the difference in frequency between upper and lower kHz QPOs in 4U 1636--53 is more
or less constant across the colour-colour diagram \citep[e.g.,][]{Mendez98, Jonker02}, using the lower kHz QPOs would lead to similar results as those shown in Figure~\ref{fig:masses}, with NS masses shifted toward higher values.

Under the assumption that the kHz QPOs are generated in the accretion disc, and considering circular orbits in the equatorial plane for Kerr space-time, the only characteristic frequencies (other than the orbital frequency) that match the observed kHz QPO frequency range are the periastron precession and the vertical epicyclic frequencies. Interpreting the upper kHz QPO as the vertical epicyclic frequency and combining the inner radius estimates with the iron line findings we found results consistent with the ones reported above. On the other hands, interpreting the upper kHz QPO as the periastron precession frequency led to meaningless NS mass values (lower than 0.1 M$_\odot$).
 
The kHz QPOs may still reflect the orbital (quasi-Keplerian) frequency at a radius far from the inner edge of the disc.
A possible scenario to reconcile this idea, for instance, could be a mechanism that amplifies the orbital frequencies of matter orbiting within a narrow ring in the disc to produce the QPO. The process could be similar to the lamp-post model by \citet{Matt91}. Such mechanism, however, must be able to pick a narrow range of radii in order to reproduce the observed high QPO coherence values \citep[e.g.,][]{Barret05b}. For instance, for a 1.8 solar mass neutron star with a QPO at 800 Hz, if this is the Keplerian frequency in the disc, the putative mechanism should pick a ring of $\sim$600 m to produce a QPO with $Q = 200$. Besides generating the kHz QPOs, this mechanism should also affect other properties of the disc, such as the emissivity index or the ionisation balance, which would in turn affect the properties of the iron emission line.
  
From the behaviour of the time derivative of the frequency of the lower kHz QPO, \citet{Sanna12a} found that the kHz QPOs (both the lower and the upper) in 4U 1636--53 are consistent with the orbital frequency at the sonic radius in the accretion disc. We also note that
the frequency of the upper kHz QPO increases monotonically across the colour-colour diagram. All this lends
support to the interpretation of the kHz QPO reflecting the orbital frequency at the inner edge of the accretion disc.

Besides Doppler and relativistic effects, the iron emission line can be broadened by other processes.
For example, the broadening may be ( partially) due to Compton 
scattering in a disc corona (\citealt{Misra98}; \citealt{Misra99}; 
see also \citealt{Reynolds00}; \citealt{Ruszkowski00}; \citealt{Turner02}, and \citealt{Ng10}).
 However, \citet{Sanna13} showed that Compton broadening alone cannot explain the broad profile of the iron emission line in 4U 1636--53. \citet{Titarchuk03} argued that the red
wing of the Fe-K$\alpha$ lines is not due to Doppler/relativistic
effects, but to relativistic, optically-thick,
wide-angle (or quasi-spherical) outflows (\citealt{Laming04}; \citealt{Laurent07}, see however
\citealt{Miller04}; \citealt{Miller07}, and \citealt{Pandel08}). As explained by \citet{Titarchuk09}, in this scenario the
red wing of the iron line is formed in a strong extended wind
illuminated by the radiation emanating from the innermost part of the
accreting material.
One of the main predictions of this model is that all high-frequency
variability should be strongly suppressed.
The fact that we detected kHz QPOs and broad iron lines simultaneously
in 4U~1636--53 casts doubt on this interpretation.
Although our findings contradict this scenario, the model under
discussion has been developed for black holes, so it is not clear how
the boundary layer or the neutron star surface could change these
predictions.

Compared to other sources, the iron line in 4U 1636--53 shows unusual properties; for instance, the best-fitting inclination is $i \gtrsim 80^\circ$ \citep{Pandel08,Cackett10,Sanna13}, which is at odds with the lack of dips or eclipse in the light curve. \citet{Pandel08} proposed that the line profile could be the blend of
two (or more) lines, for example, formed at different radii in the disc, or
due to separate regions with different ionisation balance.  If this is correct, 
the total line profile would be the result of iron lines at different energies.
To proceed further with this idea would require to solve the
ionisation balance in the accretion disc where the line is formed.
\citet{Sanna13} investigated this scenario by fitting the
reflection spectrum with a self-consistent ionised reflection model, but
 they did not find any supporting evidence for this idea \citep[see also][]{Cackett10}. 

The fact that in Obs.~3 and 5 the kHz QPO frequency significantly
varied within the 20-30~ksec required to detect the iron line
suggests that the iron line profile we model may be affected by
changes of the disc during those 20-30~ksec.
If the kHz QPO frequency depends upon the inner disc radius,
the iron line profile we observe would be the average of different line
profiles, one for each value of the inner disc radius. This is independent of whether
the kHz QPO frequency reflects the orbital disc frequency, or whether 
the relation between frequency and inner radius is more complicated.
The line energy or the disc emissivity could also vary if the inner disc radius changes.
To proceed further, detailed simulations (assuming scenarios in which
only the $R_{in}$ changes with time, as well as scenarios in which all
line parameters change) are needed to test to what extent
changes in the accretion flow on timescales of $\sim$30~ksec
(approximately the time needed with present instruments
to fit the line accurately) can affect the final line profile.

\begin{figure}

\resizebox{1\columnwidth}{!}{\rotatebox{0}{\includegraphics[clip]{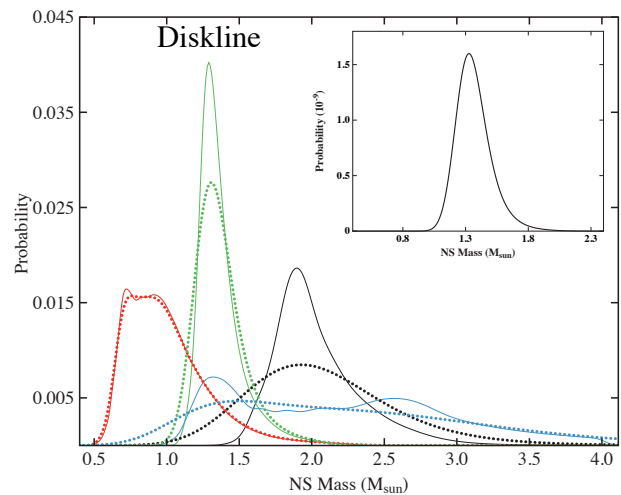}}}
\caption{Marginal probability distribution functions of the NS mass in 4U 1636--53 using the iron line model \textsc{Diskline}. Solid and dotted
lines are calculated using the error in the QPO frequency and the half-width half-maximum of the 
upper kHz QPOs, respectively. Colours are as in Figure~\ref{fig:masses}. The inset shows the joint probability distribution function for the four observations with Fe line and upper kHz QPO, using the half-width half-maximum as the error in the frequency. The area under the probability function shown in the inset represents the likelihood of measuring the masses $M_i$ of the NS mass $M_0$ under the hypothesis $H$ that one of the kHz QPOs is Keplerian, the Fe line is relativistic, and both phenomena arise from the same region of the accretion disc is valid.}
\label{fig:diskline_fwhm}
\end{figure}

A similar consideration applies to the inner radius inferred from
the kHz QPO frequency. As mentioned in Section~\ref{sec:qpo_analysis},
in Obs.~3 and 5 the frequency of the lower kHz QPO spanned a frequency range of $\sim$200 Hz during the $\sim$25 ks observation. Although, we did not directly see
the upper kHz QPO changing frequency with time, it is likely that the upper kHz QPO followed the 
lower one. If this was the case, then the full width at half maximum (FWHM)
of the upper kHz QPO observed contains information on the frequency range
covered by the QPO during the observation. To bring this information into the inner disc
radius estimates, we should use the QPO FWHM instead of the frequency error (which is relatively small)
to calculate the probability distribution function of the inner radius of the accretion disc. In Figure~\ref{fig:diskline_fwhm} we show the marginal probability distribution functions
of the NS mass inferred from the four observations combining the upper kHz QPOs and the iron lines
modelled with \textsc{diskline}. Solid and dotted lines represent the marginal probability distribution functions using the 
error in the QPO frequency and the half-width half-maximum (HWHM) as error, respectively.
In the latter case the marginal probability distribution functions of the NS mass show a broader profile, and
the range of mass values where they are consistent increases (although the overlapping area is still small). 
By combining the marginal probability distribution functions we get the mass profile (joint probability) for which,
in the 4 observations, kHz QPO and iron line estimates of the inner disc radius are consistent. 
This is shown in the inset of Figure~\ref{fig:diskline_fwhm}. The most likely value of the NS mass,
 for this specific case, ranges between $\sim$1.1 and $\sim$1.5 M$_\odot$, which is consistent with theoretical NS mass predictions \citep[e.g.,][]{Lattimer07}. 
However, it should be 
 noticed that in Figure~\ref{fig:diskline_fwhm}, two out of the four PDFs (Obs.~1 and 2) marginally overlap, therefore the final joint probability function is likely not fully representative of all observations. Statistically speaking, the overlapping area between the intersecting marginal distributions in Figure~\ref{fig:diskline_fwhm} represents the likelihood of measuring 4 values $M_i$ of the NS mass $M_0$ (assuming the mass is always the same), and the hypothesis $H$ that one of the kHz QPOs is Keplerian, the Fe line is relativistic, and both phenomena arise from the same region of the accretion disc is valid. In Table~\ref{tab:probability} we report the the NS mass values and the likelihood for the different models used to fit the Fe line profile. It is interesting to notice that the highest likelihood of the data given the model is obtained when the Fe-K$\alpha$ emission line is modelled with \textsc{diskline}.
 
\begin{table}

\resizebox{0.9\columnwidth}{!}{\begin{minipage}{\columnwidth}
\begin{tabular}{lcccc}\hline

\multicolumn{1}{c}{}  & 
\multicolumn{2}{c}{$\delta \nu$} &
\multicolumn{2}{c}{HWHM}\\\hline

\multicolumn{1}{c}{Fe line model} & 
\multicolumn{1}{c}{$M_0(M_\odot)$} &
\multicolumn{1}{c}{$\mathcal{L}$} & 
\multicolumn{1}{c}{$M_0(M_\odot)$} &
\multicolumn{1}{c}{$\mathcal{L}$} \\\hline  
 	
\textsc{Diskline} & 1.4$\pm$0.2 & 8.2\e{-3} &1.3$\pm$0.2 &1.4\e{-1} \\
\textsc{Laor} &2.6$\pm$0.6& \textless\, 1\e{-9} &3.0$\pm$1.2& 1.8\e{-6} \\
\textsc{Kyrline} a$_{*}$=0 &0.9$\pm$0.3 &\textless\, 1\e{-9} &0.9$\pm$0.3& \textless\, 1\e{-9}\\
\textsc{Kyrline} a$_{*}$=0.27&0.8$\pm$0.2 & \textless\, 1\e{-9} &0.7$\pm$0.1& 3.1\e{-3}\\ 
\textsc{Kyrline} a$_{*}$=1&0.8$\pm$0.2 & \textless\, 1\e{-9}&0.8$\pm$0.1 &1.9\e{-3}\\
\textsc{Reflection} &0.6$\pm$0.1 & 3.7\e{-3} &0.6$\pm$0.1& 2.5\e{-2}\\\hline
\end{tabular}
\end{minipage}}

\caption{Likelihood ($\mathcal{L}$) values of measuring the 4 values NS $M_i$ if the NS mass $M_0$ is always the same and under the hypothesis $H$, for different models of the Fe emission line. The two columns represent the likelihood measured from the marginal probability distribution functions of the NS mass in 4U 1636--53 calculated using the error in the QPO frequency ($\delta \nu$) and the half-width half-maximum  of the upper kHz QPOs (HWHM), respectively.}
\label{tab:probability}
\end{table}

The frequency of the kHz QPO and the source intensity (likely the mass accretion rate) are degenerate on long (longer than $\sim$ a day) time-scales \citep[``parallel tracks'',][]{Mendez99}: The same QPO frequency may appear at very different source intensities. It remains to be seen whether this phenomenon can affect some properties of the disc, such as the emissivity index or the ionisation balance, which would affect the profile of the iron line, and hence the inferred value of the inner radius of the accretion disc.

\section*{Acknowledgments}

AS wish to thank Cole Miller for interesting discussions. TB acknowledges support from ASI- INAF grant I/009/10/0 and from INAF PRIN 2012-6. AS, MM, TB and DA wish to thank ISSI for their hospitality. DA acknowledges support from the Royal Society.



\bsp

\label{lastpage}

\end{document}